\documentclass[12pt,twoside]{article}
\usepackage{amssymb}
\usepackage{amsthm}
\usepackage{graphicx}
\def\R{{\mathbb R}}
\def\N{{\mathbb N}}
\def\C{{\mathbb C}}

\def\E{{\mathbb E}}
\def\voe{{\circ\!\!\!\!\times}}
\newtheorem{thm}{Theora}[section]
\newtheorem{theo}[thm]{Theorem}

\newtheorem{cor}[thm]{Corollary}
\newtheorem{lem}[thm]{Lemma}
\newtheorem{prop}[thm]{Proposition}
\newtheorem{Def}[thm]{Definition}
\newtheorem{rem}[thm]{Remark}
\newcommand\dint{\displaystyle\int}

\newcommand\dprod{\displaystyle\prod}

\def\prend{$~~\mbox{\hfil\vrule height6pt width5pt depth-1pt}$ }
\begin{document}
\pagestyle{myheadings} \markboth{ S. H. Djah, H. Gottschalk, H. Ouerdiane }{Feynman graphs for general measures} \thispagestyle{empty}

\title{Feynman graph representation of the perturbation series
 for general functional measures}
 \author{Sidi Hamidou Djah${}^\flat$, Hanno Gottschalk${}^\sharp$ and Habib Ouerdiane${}^\flat$}
\maketitle
{\small

 \noindent ${}^\flat$: D\'epartement des Math\'ematiques, Universit\'e de Tunis El Manar

 \noindent  ${}^\sharp$: Institut f\"ur angewandte Mathematik, Rheinische Fridrich-Wilhelms-Universit\"at Bonn

\vspace{1cm}

\noindent{\bf Abstract.} A representation of the perturbation series of a general functional measure
is given in terms of generalized Feynman graphs and -rules. The graphical calculus is applied to certain functional measures of L\'evy type.
A graphical notion of Wick ordering is introduced and is compared with orthogonal decompositions of the Wiener-It\^o-Segal type.
It is also shown that the linked cluster theorem for Feynman graphs extends to generalized Feynman graphs.  We perturbatively
prove existence of the thermodynamic limit for the free energy density and the moment functions.
The results are applied to the gas of charged microscopic or mesoscopic particles -- neutral in average --
in $d=2$ dimensions generating a static field $\phi$ with quadratic energy density giving rise to a pair interaction.
 The pressure function for this system is calculated up to fourth order. We also discuss the subtraction of logarithmically divergent self-energy
terms for a gas of only one particle type by a local counterterm of first order.

\vspace{.2cm}

\noindent{\bf Key words}: {\em Feynman graphs and rules for general functional measures, Wick ordering, linked cluster theorem, free energy density,
gas of charged particles.}

\vspace{.2cm}

\noindent {\bf MSC (2000)}: \underline{82B05} 82B21, 81T15

\tableofcontents
}
\bigskip
\section{Introduction}
\label{1sec}
Let $X,Y$ be two real random variables such that their joint distribution has a unique solution of the moment
 problem and $\langle\cdot\rangle$ the expectation value.
Then $X$ and $Y$ are independent, if and only if $\langle
X^nY^m\rangle=\langle X^n\rangle\langle Y^m\rangle$ $\forall
n,m\in\N$. On the left hand side of this equation there is one
moment, but on the right hand side there is a product of two
moments. This "non-linearity of independence" expressed in terms
of moments seems harmless, but it has notable consequences
in classical statistical physics, where $X$ and $Y$ have to be
replaced by correlated random variables $\phi(x)$ and $\phi(y)$
for $x,y$ in some discrete or continuous position space, and the
independence is only asymptotic if the distance between $x$ and
$y$ goes to infinity.  The matheamtical formulationis that  the
translation group acts ergodically on the $L^2$-spaceof the underlying measure or,
with a little more physical flavour, that the statistical system under consideration is a pure phase.

 The "non-linearity" described above in many
cases of interest leads to a rather involved formulae for the
moment functions $\langle \phi(x_1)\cdots\phi(x_n)\rangle$. The
asymptotic independence can however be "linearized" by passing
through a combinatorial procedure to truncated moment
functions that fulfill $\langle
\phi(x_1)\cdots\phi(x_n)\rangle^T\to 0$ if the separation of the
arguments $x_1,\ldots,x_n$ becomes large.

This basic principle is mostly used in calculations, where
the asymptotic independence is decisive, like in practically all
problems connected with the thermodynamic (TD) limit. In
particular this applies to perturbative expansions, where often a
sufficiently fast decrease of the truncated functions is
all what one needs to carry out the TD limit order by order and to
calculate low orders explicitly. Quite often, it is convenient to
use graphs to keep track of all the terms that appear in the
expansions. A number of excellent textbooks are available on this
by now classical topic, see \cite{Ba,GJ,Ri,Ru,Sa,Se} to cite only
a few.

In modern texts on the subject, the combinatorial structure
of these expansions has been distilled into
the notion of abstract polymer system, which is sufficiently
flexible to be applied in most classical situations, like spin
systems, systems of particles in the continuum and Euclidean
quantum field theory. The handling of this concept however depends
on the physical situation, where some insight is needed to find
out what the polymers are and what is the activity function. While this is
satisfactory from the point of view of the given application, conceptually
it is somehow less clear.

In this article, we give a perturbative high temperature expansion
for the moment functions and the free energy density of a large
class of systems of statistical physics, containing in particular
the ones named above, that is to a large extent independent of the
nature of the unperturbed system under consideration and works for
a large class of interactions.  The expansion is only based on the
elementary combinatorics of "truncation" and hence the fundamental
feature of (asymptotic) independence. The motivation mainly stems
from the Feynman graph calculus in perturbative Euclidean quantum
field theory (EQFT), see e.g. \cite{GJ,KSF,Sa},  which we
generalize from Gaussian to arbitrary functional measures using
Feynman graphs with two kinds ("empty" and "full") of vertices.
Full vertices are the known interaction vertices whereas empty
vertices with $n$ legs simply symbolize a truncated $n$-point
function.

The article is organized as follows: Basic notations are collected
in Section \ref{2sec} and the perturbation series is introduced.
In Section \ref{3sec} we develop our generalized Feynman graph
calculus, which we apply in Section \ref{4sec} to some measures of
L\'evy type that have relations to particle systems and quantum
field theory, see the references [1--5], containing Gaussian
Euclidean quantum field theory as a special case. In fact, for
this more general class of models the Feynman rules are
particularly simple, just as in the Gaussian case. In Section
\ref{5sec} we introduce a general and measure independent
definition of Wick-ordering that is based on the graphical notion
of self-contraction. It coincides with orthogonal decompositions
of the Wiener-It\^o-Segal type \cite{GJ,It,Si} if and only if the
underlying measure is Gaussian. It is also shown that
Wick-ordering removes ultra-violet divergences in $d=2$ dimensions
for a class of models \cite{AGW3} containing also certain fields
of L\'evy type. The linked cluster theorem for generalized Feynman
graphs is the topic of Section \ref{6sec}, where we give a proof
which is only based on the combinatorics of truncation. We apply
this result to prove the existence of the TD limit of the free
energy density in perturbation theory. It is rather simple to
extend the results to the TD limit of moment functions using a
Schwinger term, which is done in Section \ref{7sec}. In Section
\ref{8sec} we finally apply the results of Section \ref{6sec} to
some particle systems in the continuum -- microscopic and
mesoscopic -- where the number of graphs is very effectively
reduced.  We consider a gas of charged particles that is neutral
in average and interacts via a $\phi^2$ energy density of the
static field generated by the particles. The pressure function for
this system in $d=2$ dimensions is calculated up to 4th order.
Even though the topic of ultra-violet divergences and
renormalization to a large extent is beyond the scope of this
article, we sketch the renormalization  of the perturbation series
by a local counterterm for a gas with only one type of particle
and logarithmic self-energy divergences, which to some extent is
similar to Gaussian $\phi^4$-theory in $d=3$ dimensions.

\section{Perturbation series for general functional measures}
\label{2sec}
Let $d\in\N$ be the dimension of the underlying
space\footnote{Obviously, most of the considerations of this
article remain valid if one replaces $\R^d$ and the Lebesgue
measure $dx$ with an arbitrary metric space $X$ with a sigma
finite measure $\sigma(dx)$.}  $\R^d$ (space-time in EQFT). Let
$\nu$ be a probability measure on the measurable space $({\cal
S}',{\cal B})$, where ${\cal S}'={\cal S}'(\R^d)$ is the space of
tempered distributions and ${\cal B}={\cal B}({\cal S}')$ the
Borel $\sigma$-ring generated by the open sets of the weak
topology on ${\cal S}'$. For $F:{\cal S}'\to\R$ or $F:{\cal
S}'\to\C$ $\nu_0$-integrable, we set $\langle
F\rangle_{\nu}=\int_{{\cal S}'}F(\phi)\,d\nu(\phi)$.

In  this article, we consider the perturbation theory for a "free"
probability measure $\nu_0$ on $({\cal S}',{\cal B})$ which is
subject to the following conditions
\begin{enumerate}
\item $\nu_0$ is supported on continuous functions; \item All
moments of $\nu_0$ exist; \item $\nu_0$ is translation invariant;
\item The translations are mixing \footnote{The only invariant
functions in that space are in the equivalence class of the
multiples of  the identity function. Furthermore
$\lim_{t\to\infty}\langle F\, H_{ta}\rangle_{\nu_0}=\langle
F\rangle_{\nu_0}\langle H\rangle_{\nu_0}$ for $F,H\in L^2(\nu_0)$
$a\in\R^d\setminus\{0\}$ and $H_{a}(\phi)=H(\phi_a)$ with $\phi_a$
being the translation of $\phi\in{\cal S}'$ by $a$. } on
$L^2(\nu_0)$.
\end{enumerate}
The  first condition does not hold true for many examples, e.g.
the Euclidean free field measures of QFT. In such cases, we
tacitly understand the measure $\nu_0$ as the ultra-violet
regularized version of the measure of interest. Problems of
renormalization would arise in the perturbation series when
removing this cut-off. This  problem is well-studied in EQFT,
where $\nu_0$ is Gaussian. An investigation of renormalization in
the general, not necessarily Gaussian, case would be of interest
but is beyond the scope of this work, see however Sections \ref{5sec}
 and \ref{8sec} for some first steps. Property no. 2 is an
obvious prerequisite for doing perturbation theory w.r.t.
polynomial interactions. The remaining properties 3. and 4.
technically only become important when discussing thermodynamic
limits (removing IR-cut-offs). But they are the main justification
for our graphical approach in the next section and that is why we
adopt them from the very beginning.

Let $v(\phi)=\sum_{p=0}^{\bar p}\lambda_p\, \phi^p$  be a
polynomial with $\bar p$ even and $\lambda_{\bar p}>0$, $\Lambda$
a bounded measurable set in $\R^d$ and $\phi$ a function from the
support of $\nu_0$. We define
\begin{equation}
\label{2.1eqa}
V_\Lambda(\phi)=\int_{\Lambda}v(\phi)\, dy\, .
\end{equation}
For $\phi\in{\cal S}'\setminus{\rm supp}\nu_0$ we set $V_\Lambda(\phi)=|\Lambda|\,v(0)$ with $|\Lambda|$ the Lebesgue volume of $\Lambda$.
\begin{lem}
\label{2.1lem}
$V_\Lambda:{\cal S}'\to\R$ is measurable.
\end{lem}
\noindent{\bf Proof.} Note that ${\rm supp}\nu_0$ by  definition
is a measurable set. For $y\in \Lambda$ and $\phi\in{\cal S}'$
define a map $e_y:{\cal S}'\to\R$ by setting $e_y(\phi)=1_{{\rm
supp}\nu_0}(\phi)\phi(y)$. Then $e_y$ is measurable as a pointwise
limit limit of the measurable expressions $1_{{\rm
supp}\nu_0}(\phi)\langle\delta^\epsilon_y,\phi\rangle$ where
$\delta_y^\epsilon$ is an approximation of the Dirac measure in
$y$ by $C^\infty_0(\R^d)$ test functions. Here we needed Condition
1 to establish pointwise convergence. Now, $v(e_y(\phi))$ is
measurable in $\phi$ and continuous in $y$.  The integral
(\ref{2.1eqa}) thus converges as a Riemannian sum and hence
$V_\Lambda$ is measurable as pointwise limit of measurable
functions.\prend

Later on we will feel free to replace the constants $\lambda_p$
with  continuous functions $\lambda_p(y)$, this obviously does not
affect Lemma \ref{2.1lem}.

The interacting measure $\nu_\Lambda$ is defined by
\begin{equation}
\label{2.2eqa}
d\nu_\Lambda(\phi)=Z_\Lambda^{-1}\,e^{-V_\Lambda(\phi)}\,d\nu_0(\phi)\, ,~~~~Z_\Lambda=Z(\Lambda,\lambda_0,\ldots,\lambda_{\bar p})=\left\langle e^{-V_\Lambda}\right\rangle_{\nu_0}.
\end{equation}
In this work, we perturbatively solve the following problems
\begin{enumerate}
\item Calculate the moments $Z_\Lambda\langle\phi(x_1)\cdots\phi(x_n)\rangle_{\nu_{\Lambda}}$ of the non normalized measure $Z_\Lambda\nu_\Lambda$. In particular, for $n=0$, we calculate the sum over states
$Z_\Lambda$;
\item Calculate the free energy density $f_\Lambda=\log Z_\Lambda/|\Lambda|$.
\item Calculate the moments $\langle\phi(x_1)\cdots\phi(x_n)\rangle_{\nu_\Lambda}$ of the interacting measure $\nu_\Lambda$;
\item Remove the infra-red cut-off $\Lambda$ for the free energy density and the moments of $\nu_\Lambda$.
\end{enumerate}
The term perturbatively means that we first expand into powers of $V_\Lambda$. Take e.g. problem no. 1:
\begin{eqnarray}
\label{2.3eqa}
Z_\Lambda\left\langle\phi(x_1)\cdots\phi(x_n)\right\rangle_{\nu_\Lambda}&=&\sum_{m=0}^\infty {(-1)^m\over m!}\left\langle \phi(x_1)\cdots\phi(x_n)V_\Lambda^m\right\rangle_{\nu_0}\nonumber\\
&=&\sum_{m=0}^\infty {(-1)^m\over m!}\sum_{p_1,\ldots,p_m=0}^{\bar p}\int_{\Lambda^m}\lambda_{p_1}\cdots\lambda_{p_m}\langle\phi(x_1)\cdots\phi(x_n)\nonumber\\
&&~~~~~~~~~~~~~~~~~~~~~~~~~~\times~~\phi^{p_1}(y_1)\cdots\phi^{p_m}(y_m)\rangle_{\nu_0}\, dy_1\cdots dy_m\nonumber\\
\end{eqnarray}
The first  identity in (\ref{2.3eqa}) has to be understood in the
sense of formal power series in the coupling parameters
$\lambda_1,\ldots,\lambda_p$. For many measures of interest, the
right hand side of (\ref{2.3eqa}) does not converge but (for $\Lambda\subseteq \R^d$ fixed) only gives
an asymptotic series, cf. the Lemma \ref{2.2lem} below. The second
identity is due to Fubini's lemma making use of
conditions\footnote{The technical formulation of condition 1) and
2) should include that $\phi(x_1)\cdots\phi(x_n)$ are
$L^1(\nu_0)$-integrable for all $x_1,\ldots,x_n\in\R^d$ and that
the moments of $\nu_0$ are continuous in $x_1,\ldots,x_n$.} 1) and
2) on $\nu_0$.

\begin{lem}
\label{2.2lem} Let $X,V\in \cap_{q\geq 1}L^q(\nu_0)$ with $V$
bounded from below.  Then $\langle Xe^{-\lambda V}\rangle_{\nu_0}$
at $\lambda=0$ is infinitely differentiable from the right. Hence,
the Taylor series expansion exists at $\lambda=0$ (but is not
necessarily analytic at that point).
\end{lem}
\noindent {\bf Proof.} As $\langle X\,e^{-\lambda
V}\rangle_{\nu_0}$ is right differential  at $\lambda=0$ if and
only if $e^{-\lambda c}\langle Xe^{-\lambda
V}\rangle_{\nu_0}=\langle Xe^{-\lambda (V+c)}\rangle_{\nu_0}$ is
differentiable from the right, we can assume $V$ to be
nonnegative. Then $|(e^{-\lambda V}-1)/\lambda| \leq V$ for
$\lambda >0$ and the differential quotient can be done inside the
expectation bracket by Lebesgue theorem. For $\lambda\geq 0$ the
right derivative is $\langle XV\,e^{-\lambda V}\rangle_{\nu_0}$
and now the argument can be iterated as $XV\in \cap_{q\geq 1}
L^q(\nu_0)$.\prend

To evaluate the perturbation series, one has to  calculate the
$m$-th summand on the right hand side of Eq. (\ref{2.3eqa}). It
obviously only depends on the moments of the free measure $\nu_0$.
One can argue that for a ergodic measure the truncated moment
functions (to be defined below) are more "elementary" than the
moments themselves and there are interesting examples that
illustrate this point of view. It is therefore desirable, to
expand (\ref{2.3eqa}) into such "elementary" objects. The
combinatorial book-keeping of this expansion will be done
utilizing a generalized kind of Feynman graphs.

\section{A graphical representation of the combinatorics of truncation}
\label{3sec}
The calculus of generalized Feynman graphs that is being proposed here is a device to decompose the
moments in the perturbation series
\begin{equation}
\label{3.1eqa}
\sum_{p_1,\ldots,p_m=0}^{\bar p}\int_{\Lambda^m}\lambda_{p_1}\cdots\lambda_{p_m}\langle\phi(x_1)\cdots\phi(x_n)\phi^{p_1}(y_1)\cdots\phi^{p_m}(y_m)\rangle_{\nu_0}\, dy_1\cdots dy_m\nonumber\\
\end{equation}
into truncated\footnote{Depending on the background, truncated
moments are also called "cummulants", "Ursell functions" or
"connected Greens functions". The notion "truncated moment
functions" or equivalently "truncated Schwinger functions" stems
from quantum field theory, which here is the main source of
inspiration. In the literature, the term "truncated Greens
function" often is used for a evaluation of a graphic object with
"amputatded" outer legs. Such objects in this text shall be called
"amputaded" (truncated) moment functions.} objects. In order to
explain this point of view, let us recall some well-known facts.
For a measure $\nu_0$ that is mixing, we have the cluster property
for moments
\begin{equation}
\label{3.2eqa}
\lim_{t\to\infty}\langle\phi(x_1)\cdots\phi(x_j)\phi(x_{j+1}+at)\cdots\phi(x_n+at)\rangle_{\nu_0}=\langle\phi(x_1)\cdots\phi(x_j)\rangle_{\nu_0}\langle\phi(x_{j+1})\cdots\phi(x_n)\rangle_{\nu_0}
\end{equation}
and we note that this equation  formally is non-linear in $\nu_0$.
Passing from ordinary moment functions to truncated (connected)
moment functions just provides a linearization of this equation.
As objects fulfilling a linear equation often are more simple than
objects that fulfill nonlinear constraints, it is a reasonable
step to decompose (\ref{3.1eqa}) into such truncated objects. Of
course, these general considerations have to prove useful when
dealing with concrete examples.

Let us now pass on to the technicalities. Let $J\subseteq \N$ be a
finite set. The collection of all partitions of $J$ is denoted by
${\cal P}(J)$. A partition is a decomposition of $J$ into
disjoint, nonempty subsets, i.e. $I\in{\cal P}(J)$
$\Leftrightarrow$ $\exists k\in\N$, $I=\{I_1,\ldots,I_k\}$,
$I_j\subseteq S$, $I_j\cap I_l=\emptyset$ $\forall$  $1\leq
j<l\leq k$, $\cup_{l=1}^kI_l=J$.

\begin{Def}
\label{3.1def}
{\rm
Let $J\subseteq \N$ be a finite set and $\langle J\rangle_{\nu_0}=\langle \prod_{j\in J}\phi(x_j)\rangle_{\nu_0}$ be the collection of moment functions
of $\nu_0.$ The truncated moment functions ${\langle
J\rangle}^T_{\nu_0}=\langle \prod_{j\in J}\phi(x_j)\rangle_{\nu_0}^T$ of $\nu_0$ are recursively defined
(in $\sharp J \in \N$) as follows:
\begin{equation}
\label{3.3eqa}
\langle
 J \rangle_{\nu_0}=\sum_{I \in {\cal P}(J)\atop I=\{I_1,\ldots,I_k\}}
\prod_{l=1}^k{\langle I_l\rangle}^T_{\nu_0}
\end{equation}
 }
\end{Def}

\noindent Also, we sometimes identify  $J\subseteq \N$ with the
random variable $\prod_{j\in J}\phi(x_j)$. It is well known that
\begin{itemize}
\item[F1.] The truncated moment functions are symmetric under
permutation of their arguments; \item[F2.] $(\ref{3.2eqa})$
$\Leftrightarrow$  $\lim_{t\to\infty}\langle
\phi(x_1)\cdots\phi(x_j)\phi(x_{j+1}+at)\cdots\phi(x_n+at)\rangle_{\nu_0}^T=0$
$\forall n,j\in \N$.
\end{itemize}
\noindent Hence, by F2), truncation in fact "linearizes" (5).

Obviously now one can expand  the moment
$\langle\phi(x_1)\cdots\phi(x_n)\phi^{p_1}(y_1)\cdots\phi^{p_m}(y_m)\rangle_{\nu_0}$
in (\ref{3.1eqa}) into truncated objects. To illustrate, how this
allows the passage to generalized Feynman graphs, let us consider
a two point function in second order $\phi^4$-perturbation theory,
i.e. take in (\ref{3.1eqa}) $n=m=2$ and $p_1=p_2=4$ and expand
into truncated objects. If we consider one partition,  see e.g.
the one in Fig. 1, we obtain a graph as follows: We replace all
sets in the partition, symbolized in Fig. 1 by
\hspace{.45cm}\begin{picture}(0,0)\put(0,.15){\oval(1,.3)}\end{picture}\hspace{.5cm},
with a new type of vertex "$\,\circ\,$" that is connected through
edges with all points in that set. This is just a more handy
symbol for the same thing. One then obtains the graph in Fig. 1.

We now formalize the considerations  of the above example. A graph
is a geometrical object which consists of vertices, i.e points in
$\R^d,$ which can be of different types (in our case: inner/outer,
full/ empty, cf. Table 1), and non-directed edges, i.e lines
connecting exactly two vertices (intersections of lines are
ignored). We use the term "leg" for the part of the edge meeting
the vertex, see Fig. 2. A special kind of graphs -- generalized
Feynman graphs -- occur can be associated with the expansion of
(\ref{3.1eqa}):

\begin{Def}
\label{3.2def}
{\rm

Let $n,m,p_1,\ldots,p_m\in\N_0$ be fixed. A generalized  $n$-point
Feynman graph with $m$ interaction vertices of type
$p_1,\ldots,p_m$ is a graph with $n$ outer full vertices $\times$,
$m$ inner full vertices $\bullet$ with $p_j$ the num-

\hspace{-.6cm}\parbox{8cm}{ ber of edges connected to the $j$-th
inner full vertex and  an arbitrary number of empty inner vertices
$\circ$ with an arbitrary number of edges such that each edge is
connected with exactly one vertex of full and one vertex of empty
type. By definition, full vertices are } \hspace{.5cm}
\parbox{7cm}{

\begin{center}
\begin{tabular}{||l|c|c||}
\hline\hline
& Full & Empty  \\ \hline
Inner &  $\bullet$ & $\circ$ \\\hline
Outer & $\times$ & $\voe$\\
\hline\hline
\end{tabular}
\vspace{.2cm}

{\footnotesize {\bf Table 1:} Different types of vertices.}
\end{center}
}

\noindent distinguishable and have distinguishable legs whereas
empty vertices are non  distinguishable and have non
distinguishable legs\footnote{More formally: An empty vertex with
non-distinguishable legs is a point in $\R^d$. A full vertex with
$p$ legs is given by the elements $\{(y,1),\ldots, (y,p)\}$ where
$y\in\R^d$ is the point associated to that vertex and $(y,j)$ are
the legs, $j=1,\ldots,p$. Let ${\cal M}_1$ be the collection of
all empty vertices and all legs of full vertices. Let ${\cal M}_2$
be the set of all non ordered pairs of ${\cal M}_1$. A graph is a
subset of ${\cal M}_2$. A generalized Feynman graph is a graph
such that each pair in the graph consists of one point (empty
vertex) and one leg of a full vertex.}${}^,$\footnote{Note that
empty outer vertices will not be needed in this work. They are
however useful in connection with generalized renormalization
group equations where the flow can be expressed in terms of
amputated moment functions and graphs, see \cite{Go}.}, cf. Fig.
2. }

\end{Def}

\begin{figure}
\begin{center}
\begin{picture}(8,3)
\put(1,.9){$\times$}
\put(6.7,.9){$\times$}
\put(3,1.1){\circle*{.08}}
\put(3,.9){\circle*{.08}}
\put(3.4,1.1){\circle*{.08}}
\put(3.4,.9){\circle*{.08}}
\put(4.6,1.1){\circle*{.08}}
\put(4.6,.9){\circle*{.08}}
\put(5,1.1){\circle*{.08}}
\put(5,.9){\circle*{.08}}
\thinlines
\put(2,1){\oval(2.2,.5)}
\put(6,1){\oval(2.2,.5)}
\put(4,1.125){\oval(1.5,.2)}
\put(4,.875){\oval(1.5,.2)}
\put(1.9,.85){$I_1$}
\put(3.9,1.35){$ I_2$}
\put(3.9,.4){$I_3$}
\put(5.9,.85){$I_4$}
\put(3.2,1){\circle{.8}}
\put(4.8,1){\circle{.8}}
\put(1.1,.4){$x_1$}
\put(6.8,.4){$x_2$}
\put(2.6,.1){$J_1/y_1$}
\put(4.4,.1){$J_2/y_2$}

\thicklines
\put(1,2.5){$\times$}
\put(6.7,2.5){$\times$}
\put(3.05,2.5){$\bullet$}
\put(4.85,2.5){$\bullet$}
\put(2.03,2.5){$\circ$}
\put(5.83,2.5){$\circ$}
\put(3.94,2){$\circ$}
\put(3.94,2.9){$\circ$}
\put(1.15,2.6){\line(1,0){.9}}
\put(6,2.6){\line(1,0){.85}}
\bezier{100}(2.2,2.65)(2.6,2.95)(3.1,2.65)
\bezier{100}(2.2,2.55)(2.6,2.2)(3.1,2.55)
\bezier{100}(3.15,2.55)(3.6,2.15)(3.95,2.1)
\bezier{100}(3.15,2.65)(3.6,2.95)(3.95,3)
\bezier{100}(4.13,3)(4.5,2.95)(4.9,2.65)
\bezier{100}(4.13,2.1)(4.55,2.15)(4.9,2.55)
\bezier{100}(5,2.55)(5.4,2.15)(5.85,2.55)
\bezier{100}(5,2.65)(5.4,2.95)(5.85,2.65)
\end{picture}
\end{center}
\caption{\footnotesize A partition $I=\{I_1,\ldots,I_4\}$ of $n=2$
outer points $x_1,x_2$ and $8=2\times 4$ inner points
corresponding to one term in the 2nd order perturbation  theory of
the two point function in $\phi^4$ theory. The first "vertex set"
$J_1$ is the set of the first four inner points which take the
value $y_1$ and $J_2$ the set of the remaining four inner points
that take the value the value $y_2$. Above, the corresponding
generalized Feynman graph is displayed.}
\end{figure}
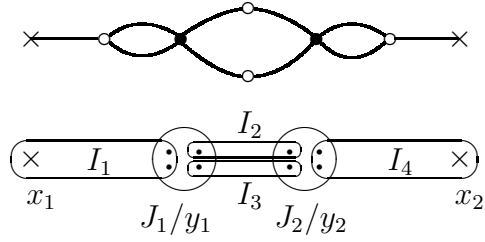

Let $n,m,p_1,\ldots,p_m\in \N_0$ and $X, J_1,\ldots, J_m\subset
\N$ disjoint  sets be given s.t. $\sharp X=n$, $\sharp
J_1=p_1,\ldots,\sharp J_m=p_m$. Then we can construct a one to
correspondence between ${\cal P}(X\cup\bigcup_{l=1}^mJ_l)$ and the
Feynman graphs with $n$ full outer vertices and $m$ full inner
vertices of type $p_1,\ldots,p_m$ that is given in the following
way: Pick an arbitrary (but fixed) bijection between the
distinguishable outer points and $X$. Pick also bijections of the
legs of the $j$-th vertex with $p_j$ (distinguishable) edges and
$J_j$, $j=1,\ldots,m$. Let $G$ be a graph as described in
Definition \ref{3.2def}. Suppose that there are $k$ empty inner
vertices in the graph. Give an arbitrary number $l=1,\ldots, k$ to
each inner vertex. For the $l$-th empty inner vertex let $I_l$ be
the set of all points in $X\cup\bigcup_{j=1}^mJ_j$ that correspond
under the to the given bijections with the edges connected to that
empty vertex. The the partition associated to $G$ is given by
$I=\{I_1,\ldots,I_k\}$.

Conversely,  let $I=\{I_1,\ldots,I_k\}\in{\cal
P}(X\cup\bigcup_{l=1}^mJ_l)$ be given. Draw $n$ outer full
vertices, $m$ inner full vertices with $p_1,\ldots,p_m$ legs and
$k$ inner empty vertices with $\sharp I_1,\ldots,\sharp I_k$ legs.
Connect the legs of $l$-th inner empty vertex with all the legs of
inner full vertices or outer full vertices corresponding -- under
the fixed bijections -- to the points in $I_l$, $l=1,\ldots,k$.
The result obviously is a generalized Feynman graph. Hence one
obtains a mapping from ${\cal P}(X\cup\bigcup_{l=1}^mJ_l)$ to the
generalized Feynman graphs as described in Definition
\ref{3.2def}. The inverse of this mapping clearly is the mapping
described in the previous paragraph and vice versa. We have thus
deived

\begin{lem}
\label{3.1lem} Let  $n,m,p_1,\ldots,p_m\in\N_0$ and $J_1,\ldots
J_m$ as above. Then there exists a one to one correspondence
between ${\cal P}(X\cup\bigcup_{l=1}^mJ_l)$ and the generalized
Feynman graphs as described in Definition \ref{3.2def}.
\end{lem}

Given the interaction  polynomial $v(\phi)=\sum_{p=0}^{\bar
p}\lambda_p\,\phi^p$, let ${\cal F}(n,m)={\cal F}(n,m,v)$ be the
collection of all generalized Feynman graphs with $n$ outer full
vertices and $m$ inner full vertices such that each inner full
vertex has a number $p$ of edges such that $1\leq p\leq\bar p$ and
$\lambda_p\not=0$. The following definition that assigns a
numerical value to each Feynman graph in the physical literature
goes under the name "Feynman rules":
\begin{figure}[t]
\begin{center}
\begin{picture}(12.2,2)
\thicklines

\put(.3,.5){\line(1,0){.8}}
\put(1.09,.4){$\bullet$}
\put(1.1,.55){\line(-2,1){.8}}
\put(1.15,.5){\line(1,0){.8}}
\put(1,.9){$\cdots$}
\put(0,.4){$1$}
\put(2,.4){$n$}
\put(0,1.1){$2$}

\put(2.45,.6){$\not =$}

\put(3.3,.5){\line(1,0){.8}}
\put(4.09,.4){$\bullet$}
\put(4.1,.55){\line(-2,1){.8}}
\put(4.15,.5){\line(1,0){.8}}
\put(4,.9){$\cdots$}
\put(3,.4){$2$}
\put(5,.4){$n$}
\put(3,1.1){$1$}

\put(9.45,.6){$=$}

\put(7.3,.5){\line(1,0){.8}}
\put(8.09,.4){$\circ$}
\put(8.1,.55){\line(-2,1){.8}}
\put(8.25,.5){\line(1,0){.7}}
\put(8,.9){$\cdots$}
\put(7,.4){$1$}
\put(9,.4){$n$}
\put(7,1.1){$2$}

\put(10.3,.5){\line(1,0){.8}}
\put(11.09,.4){$\circ$}
\put(11.1,.55){\line(-2,1){.8}}
\put(11.25,.5){\line(1,0){.7}}
\put(11,.9){$\cdots$}
\put(10,.4){$2$}
\put(12,.4){$n$}
\put(10,1.1){$1$}

\end{picture}
\end{center}
\caption{\footnotesize Distinguishable and non-distinguishable legs}
\end{figure}
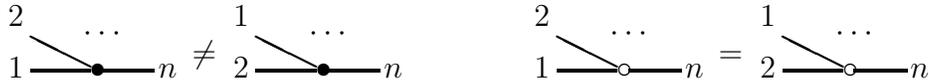

\begin{Def}
\label{3.3def} {\rm Let $G\in{\cal F}(n,m)$  and
$x_1,\ldots,x_n\in\R^d$ be given. Then the real number ${\cal
V}_\Lambda[G]={\cal V}_\Lambda[G](x_1,\ldots,x_n)$ is obtained in
the following way:
\begin{enumerate}
\item Assign the values $x_1,\ldots,x_n$ to  the outer full
vertices of the graph and assign arbitrary values $y_1,\ldots,y_m$
to the inner full vertices;
\item For each inner empty vertex
with $l$ legs multiply with a truncated $l$-point moment function with arguments
given by the full vertex points where the $l$ edges connected to
that vertex are ending;
\item Multiply with $\lambda_p$ for each
inner full vertex with $p$ legs;
\item Integrate the inner full
vertices $y_1,\ldots,y_m$ over $\Lambda$ (w.r.t. the Lebesgue
measure).
\end{enumerate}
}
\end{Def}
The value of  ${\cal V}_\Lambda[G](x_1,\ldots,x_n)$ is obviously
just the one of the term in the expansion of (\ref{3.1eqa}) into
truncated objects that corresponds to the partition associated
with $G$. Note that by F1) this value is independent of the
bijections between legs of full vertices and the sets
$J_1,\ldots,J_m$. In general, it does depend on the chosen bijection
between $X$ and the outer full vertices, this dependence however is
eliminated in sums over all generalized Feynman graphs. Combining Lemma \ref{3.1lem}, Definition
\ref{3.3def} and Equation \ref{3.1eqa} one thus gets
\begin{theo}
\label{3.1theo} The $n$-point functions  of the non-normalized
interacting measure $Z_\Lambda \nu_\Lambda$ are given in the sense
of formal power series by a sum over all generalized Feynman
graphs with $n$ exterior full points that are evaluated according
to the Feynman ruled fixed in Def. \ref{3.3def}, i.e.
\begin{equation}
\label{3.4eqa}
Z_\Lambda\langle\phi(x_1)\cdots\phi(x_n)\rangle_{\nu_\Lambda}=\sum_{m=0}^\infty {(-1)^m\over m!}\sum_{G\in{\cal F}(n,m)}{\cal V}_\Lambda[G](x_1,\ldots,x_n) \, .
\end{equation}
\end{theo}
All graphs that differ only by the labelling of full inner vertices and
edges of inner full vertices give the same value ${\cal V}_\Lambda[G]$.
The equivalence class of graphs under permutations of legs of full
vertices and full vertices is called topological generalized
Feynman graph and the perturbation series in Equation
(\ref{3.4eqa}) can equally be expressed through a sum over
topological Feynman graphs where the multiplicity factor, i.e. the
number of elements in the equivalence class, is built in into the
Feynman rules. Calculating the multiplicity in concrete cases can
be rather complicated. A first step in that direction is to make
the legs at a full interaction vertex non-distinguishable:
\begin{cor}
\label{3.1cor} If one replaces the interaction density
$v(\phi)=\sum_{p=0}^{\bar p}\lambda_p\,\phi^p$ with
$v(\phi)=\sum_{p=0}^{\bar p}\lambda_p\, \frac{\phi^p}{p!}$, the
generalized Feynman graphs and rules change in the following way :
the $\bullet$ vertices are treated like vertices with
non-distinguishable legs. When evaluating, for each inner full
vertex "$\bullet$" one has to multiply by
\begin{equation}
\label{3.5eqa}
\frac{1}{\rm \dprod_{\begin{array}{lll}\rm all ~"\circ"~
vertices\\ \rm directly\; connected\\ \rm to\; \bullet\; by\; an\;
edge\end{array}}\sharp\{edges\; from\; \bullet\; to\; \circ\}!}\,.
\end{equation}
\end{cor}
The advantage of this prescription is that treating the edges at
the interaction vertex as indistinguishable considerably reduces
the combinatorics of generalized Feynman graphs.

A further reduction of this combinatorics takes place, if certain
truncated moment functions of $\nu_0$ vanish identically. Then,
one can omit the corresponding empty vertices from the
perturbation series. A particularly interesting case arises from
the following well-known fact
\begin{itemize}
\item[F3.] All odd truncated moment functions vanish if and only if all odd moment function vanish.
\end{itemize}
\begin{cor}
\label{3.2cor} Let the measure $\nu_0$ be symmetric under the
mapping $\phi\to-\phi$, i.e. $\nu_0(A)=\nu_0(-A)$, $\forall$
$A\in{\cal B}$. Then one can omit all such generalized Feynman
graphs from the perturbation series that have an empty inner
vertex with an odd number of legs.
\end{cor}

\section{Application to certain functional measures of L\'evy type}
\label{4sec} In this section,  we give a justification to the
general procedure of Section \ref{3sec} by the means of examples.
In particular, we consider the case  where $\nu_0$ is a convoluted
generalized white noise measure in the sense of \cite{AGW1}. This
gives a unified treatment of the perturbation expansion around the
Gaussian Euclidean free field measure in QFT and the case the high
temperature expansion of classical, continuous particles in the
grand canonical ensemble, cf. \cite{AGY1,AGY2} and Section
\ref{8sec}. In the first -- Gaussian -- case, generalized Feynman
graphs and rules reduce to the classical Feynman graphs and rules.
In the more general L\'evy case, one still obtains Feynman rules
that are very close to the original ones of R. P. Feynman
\cite{GJ,Fe}. This simple observation, namely that full and empty
vertices in the Feynman rules can be treated on the same level, is
the crucial argument in favor of the generalized Feynman graph
formalism of Section \ref{3sec}.

Firstly, let us recall some well-known  technicalities: Let ${\cal
C}:{\cal S}\to\C$ with ${\cal S}$ the space of Schwartz test
functionsover $\R^d$. The Frechet derivative of ${\cal C}$ at $h\in{\cal S}$
in direction $u\in{\cal S}$ is by definition ${\partial{\cal
C}(h)\over\partial u}=\lim_{t\to 0,t\not=0}({\cal C}(h+tu)-{\cal
C}(h))/t$ provided this limit exists. The functional derivative of
${\cal C}(h)$ w.r.t. $\phi(x)$ is defined as ${\delta {\cal
C}(h)\over\delta\phi(x)}=\lim_{u\to\delta_x}{\partial{\cal
C}(h)\over\partial u}$ where $\delta_x$ is the Dirac measure of
mass one in $x$ an the convergence $u\to\delta_x$ is in the sense
of the weak topology of signed Borel measures in ${\R^d}$.

It is easy to show that the characteristic function ${\cal
C}_{\nu_0}(h)=\langle e^{i\langle h,.\rangle}\rangle_{\nu_0}$ of
the measure $\nu_0$ under the conditions 1. and 2. of Section
\ref{2sec} has functional derivatives of arbitrary order.
Obviously, ${\cal C}_{\nu_0}$ is the generating functional of the
sequence of moments of $\nu_0$, i.e.
$\langle\phi(x_1)\cdots\phi(x_n)\rangle_{\nu_0}=(-i)^n{\delta^n{\cal
C}_{\nu_0}(h)\over\delta \phi(x_1)\cdots\delta\phi(x_n)}|_{h=0}$.
Let ${\cal C}_{\nu_0}^T=\log{\cal C}_{\nu_0}$, then ${\cal
C}^T_{\nu_0}(h)$ is well defined for $h\in{\cal S}$ sufficiently
small as ${\cal C}_{\nu_0}(h)$ is continuous in $h$ and ${\cal
C}(0)=1$. Furthermore, also ${\cal C}^T_{\nu_0}$ has functional
derivatives of arbitrary order. The crucial fact needed in this
section is the basic linked cluster theorem
\begin{itemize}
\item[F4.] ${\cal C}^T_{\nu_0}$ is the generating functional of the sequence of truncated moment functions.
\end{itemize}

Minlos theorem \cite{Mi} establishes a one to one correspondence
between characteristic functionals ${\cal C}:{\cal S}\to\C$ (
positive definite normalized ${\cal C}(0)=1$ and continuous)
random fields $\eta$ indexed by ${\cal S}$ (up to equivalence in
law), cf. \cite{It}, and probability measures $\rho_0$ on $({\cal
S}',{\cal B})$ given by ${\cal C}(h)=\langle e^{\langle
h,.\rangle}\rangle_{\rho_0}=\E[e^{i\eta(h)}]$. To define a measure
$\rho_0$, it is thus sufficient to write down its characteristic
functional. Let us do this for noise (infinitely divisible and
non-correlated at a distance) measures of L\'evy type.

Let $\psi:\R\to\C$ be a L\'evy characteristic (conditionally
positive definite, normalized $\psi(0)=0$ and continuous)
\cite{BF} that is infinitely often differentiable at zero. Then,
$\psi(t)$ has the following representation
\begin{equation}
\label{4.1eqa}
\psi(t)=iat-{\sigma^2\over 2}t^2+z\int_{\R\setminus\{0\}}(e^{ist}-1)\, dr(s),
\end{equation}
where $a\in\R$, $\sigma^2,z\geq 0$ and $r$ is a probability
measure on $\R\setminus\{0\}$ that has all moments. The first term
in (\ref{4.1eqa}) is called deterministic, the second one Gaussian
part and the third one Poisson part. If $z>0$, the representation
(\ref{4.1eqa}) is unique. It is well-known, cf. Theorem 6 of
\cite{GV} p. 238, that ${\cal
C}_{\rho_0}(h)=\exp\{\int_{\R^d}\psi(h) \, dx\}$, $h\in{\cal S}$,
defines a characteristic functional.

Let $\rho_0$ be the associated measure on $({\cal S}',{\cal B})$
and $\eta$ the associated coordinate process, i.e.
$\eta(h)(\omega)=\omega(h)$ $\forall h\in{\cal S},\omega\in{\cal
S}'$. We consider the linear stochastic partial differential
equation (SPDE) $L\phi=\eta$ with $L:{\cal S}'\to{\cal S}'$ a
partial (pseudo) differential operator with constant coefficients
and with Greens function $g:\R^d\to\R$, i.e. $g*L\omega=\omega$ for
$\omega\in{\cal S}'$. As the most relevant case, we consider
$L=(-\Delta+m_0^2)^{\alpha}$ for $\alpha>0,m_0>0$ and $\Delta$ the
Laplacian on $\R^d$. Then, the solution to this SPDE $\phi=g*\eta$
exists pathwisely.  As a canonical process it is equivalent (in
distribution) to the coordinate process of the measure $\nu_0$ on
$({\cal S}',{\cal B})$ with characteristic functional
\begin{equation}
\label{4.2eqa}
{\cal C}_{\nu_0}(h)=\exp\{\int_{\R^d} \psi(g*h) \, dx \} .
\end{equation}

It is easily  verified that for $L=(-\Delta+m_0^2)^{1\over 2}$ and
$a,z=0$, $\nu_0$ is the free field measure of Euclidean QFT
(Nelson's free field measure, cf. \cite{GJ,Si}). But also in the
more general case considered here, connections with quantum field
theory can be made explicit \cite{AGW1}.

It turns out  \cite{AGW1} that the measure $\nu_0$ obtained in
this way fulfills the conditions 2. -- 3. of Section \ref{2sec},
however in general does not fulfill Condition 1. This can be seen
as a ultra-violet problem and can be removed replacing $g$ with
$g_\epsilon=g*\chi_\epsilon$ where $\chi_\epsilon\in{\cal S}$ is
an approximation of the Dirac delta distribution in zero,
$\chi_\epsilon\to\delta_0$ in ${\cal S}'$. The measure
$\nu_0^\epsilon$ with characteristic functional (\ref{4.2eqa})
where $g$ is replaced by $g_\epsilon$ then also fulfills Condition
1\footnote{$\nu^\epsilon_0$ is the image measure of $\nu_0$ under
the mapping ${\cal S}'\ni\phi\to\phi*\chi_\epsilon\in
C^\infty(\R^d)$.}. In the following we will tacitly assume that
measures $\nu_0$ are suitably ultra-violet regularized and we do
not write the superscript $\epsilon$. Some simple examples, where
the ultra-violet cut-off can be removed in the perturbation series
can be found in the Sections \ref{5sec} and \ref{8sec}. The
ultra-violet problem for the general case of convoluted L\'evy
noise has to be postponed.

Combination of F4. with (\ref{4.2eqa}) now yields
\begin{center}
\begin{picture}(14,1.5)
\thicklines
\put(.3,.5){\line(1,0){.8}}
\put(1.08,.4){$\circ$}
\put(1.1,.55){\line(-2,1){.8}}
\put(1.25,.5){\line(1,0){.7}}
\put(1,.9){$\cdots$}
\put(0,.4){$1$}
\put(2,.4){$n$}
\put(0,1.1){$2$}
\put(2.45,.6){$ =$}
\put(3,.6){$\langle\phi(x_1)\cdots\phi(x_n)\rangle_{\nu_0}^T=c_n\dint_{\R^d}g(x_1-z)\cdots g(x_n-z)\, dz\, ,$}
\put(14.25,.6){$(11)$}
\end{picture}
\setcounter{equation}{11}
\end{center}
where
\begin{equation}
\label{4.3eqa}
c_n=(-i)^n{d^n\psi(t)\over dt^n}|_{t=0}=\delta_{n,1}\, a+\delta_{n,2}\,\sigma^2+z\int_{\R\setminus\{0\}}s^n\, dr(s)\,,
\end{equation}
$\delta_{n,n'}$ being the Kronecker symbol. Note that the property
F2. obviously holds for the truncated moments (11) for $g$ of
sufficiently fast decay. From equation (11) one now obtains the
Feynman rules for convoluted L\'evy type noise:

\begin{theo}
\label{4.1theo}
Let $G\in{\cal F}(n,m)$ and $x_1,\ldots,x_n\in\R^d$ be given. For the case of a convoluted L\'evy noise measure $\nu_0$ the
value of ${\cal V}_\Lambda[G]={\cal V}_\Lambda[G](x_1,\ldots,x_n)$ can be calculated as follows:
\begin{enumerate}
\item Assign the values $x_1,\ldots,x_n$  to the outer full
vertices of the graph and assign arbitrary values $y_1,\ldots,y_m$
to the inner full vertices and $z_1,\ldots,z_k$ to the inner empty
vertices where $k$ is the number of such vertices; \item For each
edge in the graph going from a full vertex $x_j$ or $y_j$ to an
empty vertex $z_q$ multiply with the "propagator function"
$g(x_j-z_q)$ and $g(y_j-z_q)$, respectively; \item For each inner
empty vertex with $l$ legs multiply with $c_l$; \item Multiply
with $\lambda_p$ for each inner full vertex with $p$ legs; \item
Integrate over all inner vertices $y_1,\ldots,y_m$ and
$z_1,\ldots,z_k$ (w.r.t. the Lebesgue measure) -- full vertices
are being integrated over $\Lambda$ and empty ones over $\R^d$.
\end{enumerate}
\end{theo}
In Theorem \ref{4.1theo}, the constants $c_l$, $l\in\N$,  take the r\^ole of coupling
constants of empty vertices. Hence empty and full inner vertices in
the Feynman rules are treated on the same level -- at least in the
thermodynamic limit $\Lambda\nearrow\R^d$.

Let us consider the centered Gaussian  case $a=z=0$ as a special
case. Then, $c_l=0$ for $l=1$ and $l\geq 3$, cf. (\ref{4.2eqa}).
Hence all graphs containing empty vertices with a number of legs
not equal to two give a zero contribution. The remaining two
legged empty vertices $1~\mbox{\bf
-}\hspace{-.15cm}\circ\hspace{-.13cm}\mbox{\bf
-}~2=c_2g*g(x_1-x_2)$ can be identified with a straight line of a
new type. Hence one obtains the classical Feynman graphs and
-rules as a special case:

\begin{cor}
\label{4.1cor} In the case where $\nu_0$ is a centered  Gaussian
measure, i.e. $a=z=0$, there exists a one to one correspondence
between the generalized Feynman graphs that give non-zero
contributions in the Feynman rules, i.e. that contain only
two-legged empty vertices, and the classical Feynman graphs, cf.
Fig. 3.

Furthermore, the generalized Feynman rules of Theorem
\ref{4.1theo} with propagator function $g$ applied to a
generalized Feynman graph and the classical Feynman rules applied
to the corresponding classical Feynman graph with propagator
$g_1=c_2\, g*g$ give the same result.
\end{cor}

\begin{rem}
\label{4.1rem} {\rm In \cite{AGW1} the  moment functions of
convoluted L\'evy noise have been analytically continued to vacuum
expectation values (Wightman functions) of a local, relativistic
QFT  that fulfill all Wightman axioms \cite{SW} except positivity.
Thus, for the non-interacting case, there is a correspondence
between convoluted L\'evy noise and a relativistic, local quantum
field theory with indefinite metric \cite{AGW2}.

It is an interesting speculation that this correspondence exists
also in the interacting case. We note that by Theorem
\ref{4.1theo} all Feynman graphs correspond to a Feynman graph in
some Gaussian theory with modified propagator and interaction
structure. The contribution to the Wightman function that
corresponds to such a graph, is known at least in principle, i.e.
in non-renormalized form \cite{Os,Ste1,Ste2}. It is natural to
conjecture, that the analytic continuation of the function
corresponding to the Euclidean Feynman graph is given by that part
of the Wightman function. One can show that the analytic
continuation obtained in \cite{AGW1} is equal to the expression in
\cite{Os,Ste1,Ste2} obtained for the sectorized star graph. Hence
this conjecture holds for star graphs (11).

If it would be true in general,  one would obtain a perturbative
correspondence of convoluted L\'evy noise with local, polynomial
interactions and local, relativistic Quantum fields with
indefinite metric. In particular, the expectation values of
products of the static field of a ensemble of interacting
particles, see \cite{AGY1,AGY2} and Section \ref{8sec} for further
explanations, would have a relativistic, local Wightman function
as its counterpart, as already conjectured in \cite{AGY2}. The
above argument above gives new evidence in favor of this
conjecture. }\prend
\end{rem}
\begin{figure}
\begin{center}
\begin{picture}(12,1.5)

\thicklines
\put(.5,.49){\line(1,0){.5}}
\put(0.98,.385){$\circ$}
\put(1.15,.49){\line(1,0){.5}}
\put(2,.385){$=$}
\bezier{7}(2.5,.5)(2.85,.5)(3.2,.5)
\put(.65,.7){$g$}
\put(1.3,.7){$g$}
\put(2.75,.7){$g_1$}
\put(4,.385){$\Rightarrow$}
\put(5.05,.385){$\times$}
\put(5.21,.49){\line(1,0){.25}}
\put(5.44,.385){$\circ$}
\put(5.61,.49){\line(1,0){.25}}
\put(5.84,.385){$\bullet$}
\put(6.01,.49){\line(1,0){.25}}
\put(6.24,.385){$\circ$}
\put(6.41,.49){\line(1,0){.25}}
\put(6.64,.385){$\bullet$}
\put(6.82,.49){\line(1,0){.25}}
\put(7.05,.385){$\circ$}
\put(7.22,.49){\line(1,0){.28}}
\put(7.34,.385){$\times$}
\put(6.24,.085){$\circ$}
\put(6.24,.685){$\circ$}
\bezier{100}(5.9,.55)(6.05,.75)(6.25,.8)
\bezier{100}(5.9,.45)(6.05,.25)(6.25,.2)
\bezier{100}(6.42,.8)(6.62,.75)(6.75,.55)
\bezier{100}(6.42,.2)(6.62,.25)(6.75,.45)
\put(8,.385){$=$}
\put(8.79,.385){$\times$}
\put(9.6,.385){$\bullet$}
\put(10.4,.385){$\bullet$}
\put(11.04,.385){$\times$}
\bezier{6}(8.95,.49)(9.3,.49)(9.65,.49)
\bezier{6}(9.7,.49)(10.05,.49)(10.4,.49)
\bezier{6}(10.5,.49)(10.85,.49)(11.2,.49)
\bezier{7}(9.7,.55)(10.125,1)(10.45,.53)
\bezier{7}(9.7,.45)(10.125,0)(10.45,.45)
\end{picture}
\end{center}
\caption{\footnotesize Identification  of a generalized Feynman
graph with propagator $g$ with a classical Feynman graph with
propagator $g_1$ in the case of a Gaussian functional measure.}
\end{figure}
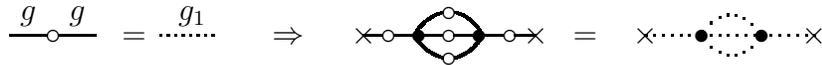

\section{Wick--ordering vs Wiener-It\^o-Segal chaos decomposition}
\label{5sec} In this section we give the notion of  Wick ordering
for a general functional measure. When removing ultra-violet
cut-offs, take e.g. $\epsilon\searrow 0$ in the examples given in
Section \ref{4sec}, some graphs in the expansion introduced in
Section \ref{3sec} will diverge. The reason for these divergences
is that the truncated moment functions of a non-uv-regularized
measure have singularities when two or more of its arguments
coincide. The worst of these cases, i.e. the one with the
strongest divergences, certainly is the one when all of the
arguments of a truncated $n$-point moment function coincide and a
term $\sim\langle\phi^n(y)\rangle^T_{\nu_0}$ occurs in the Feynman
rules. In graphical terms, this situation corresponds to a
self-contraction, i.e. to the case where all $n$ legs of an empty
vertex are connected to one and the same inner full vertex, cf.
Fig. 4. Wick ordering -- as it is understood here -- removes
graphs with self-contractions from the perturbation series. In the
general case, this does not yet render the perturbation series
finite, and more sophisticated procedures of renormalizataion have
to be applied to achieve that. Some remarkable exceptions --
Gaussian and not -- in $d=2$ dimensions will be discussed at the
end of this section. We also clarify the relation of Wick ordering
in the given sense and the decomposition of $L^2(\nu_0)$ by means
of orthogonal polynomials, e.g. of Hermite \cite{GJ,Si} or Charlier \cite{IK,Pr} type that
goes under the name of Wiener-It\^o-Segal chaos decomposition.

Let $X\subseteq \N$  a set of numbers and $Y\in L^2(\nu_0)$ a
random variable. When considering $X\cup\{Y\}$ as a collection of
$\sharp X+1$ objects, we can use Definition \ref{3.1def} to make
sense of $\langle X\, Y\rangle^{(T)}_{\nu_0}=\langle \prod_{j\in
X} \phi(x_j)Y\rangle^{(T)}_{\nu_0}$. Here the symbol $(T)$ is
being used instead of $T$ in order to symbolize that the random
variable $Y$ in the combinatorics of Def. \ref{3.1def} is treated
as one object in order to avoid ambiguities if e.g.
$Y=\phi(y_1)\cdots \phi(y_n)$. The field entries from
$X=\prod_{j\in X}\phi(x_j)$ combinatorially are treated as
distinct objects. We are now looking for another random variable,
denoted by $:X:=:X:_{\nu_0}$, that has the same $L^2(\nu_0)$ inner
product with an arbitrary $L^2(\nu_0)$ random variable  as $X$
with the exception that there are no self-contractions in $X$,
i.e.
\begin{equation}
\label{5.1eqa}
\langle :X:Y\rangle_{\nu_0}=\langle X\, Y\rangle^{(T)}_{\nu_0} ~~\forall Y\in L^2(\nu_0).
\end{equation}

\begin{figure}
\begin{center}
\begin{picture}(13,1.5)

\thicklines
\put(1.99,.5){\circle*{.15}}
\put(1.99,1.1){\circle{.15}}
\put(1.99,1.03){\line(0,-1){.5}}
\put(1.5,.51){\line(1,0){.5}}
\put(2.05,.51){\line(1,0){.5}}
\put(1.99,.49){\line(0,-1){.5}}

\put(5,.5){\circle*{.15}}
\put(5,1.1){\circle{.15}}
\bezier{100}(4.92,1.08)(4.7,.73)(4.93,.52)
\bezier{100}(5.07,1.08)(5.27,.73)(5.06,.52)
\put(4.95,.48){\line(-1,-1){.4}}
\put(5.05,.48){\line(1,-1){.4}}

\put(8,.5){\circle*{.15}}
\put(8,1.1){\circle{.15}}
\bezier{100}(7.93,1.08)(7.7,.73)(7.94,.52)
\bezier{100}(8.07,1.08)(8.29,.73)(8.05,.52)
\put(8,.49){\line(0,-1){.5}}
\put(8,1.03){\line(0,-1){.5}}

\put(10.98,.08){\circle*{.15}}
\put(10.98,1.1){\circle{.15}}
\bezier{100}(10.9,1.06)(10.5,.53)(10.9,.12)
\bezier{100}(11.05,1.06)(11.45,.53)(11.05,.12)
\bezier{100}(10.92,1.06)(10.8,.53)(10.92,.12)
\bezier{100}(11.03,1.06)(11.15,.53)(11.03,.12)

\end{picture}
\end{center}
\caption{\footnotesize Self-contractions at a $\phi^4$ interaction vertex.}
\end{figure}
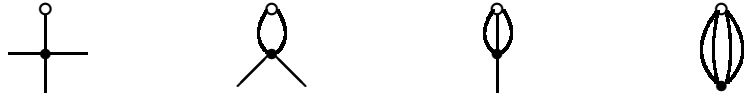

\noindent By Def. \ref{3.1def} one has
\begin{eqnarray}
\label{5.2eqa}
\langle X\, Y\rangle_{\nu_0}&=&\sum_{I\in{\cal P}(X\cup \{Y\})\atop I=\{I_1,\ldots,I_k\}}\prod_{l=1}^k\langle I_k\rangle_{\nu_0}^{(T)}\nonumber\\
&=&\sum_{I\in{\cal P}(X)\atop I=\{I_1,\ldots,I_k\}}\sum_{j=1}^k\langle I_j Y\rangle_{\nu_0}^{(T)}\prod_{l=1\atop l\not=j}^k\langle I_l\rangle^T_{\nu_0}
+\langle Y\rangle_{\nu_0}\sum_{I\in{\cal P}(X)\atop I=\{I_1,\ldots I_k\}}\prod_{l=1}^k\langle I_l\rangle ^T_{\nu_0}~,
\end{eqnarray}
We note that only the  $k=1$ term in the first sum of
(\ref{5.2eqa}) appears on the right hand side of (\ref{5.1eqa}).
Hence, using linearity in $Y$, one can show that the following is
the only solution to (\ref{5.1eqa}):

\begin{Def}
\label{5.1def} {\rm For $X\subseteq \N$  with $\sharp X=1$ let
$:X:=X-\langle X\rangle_{\nu_0}$. Let now $X\subseteq\N$ with
$\sharp X>1$ and suppose that $:J:$ is already defined for
$J\subseteq \N$ with $\sharp J<\sharp X$. Then
\begin{equation}
\label{5.3eqa}
:X:=X-\langle X\rangle_{\nu_0}-\sum_{I\in{\cal P}(X)\atop I=\{I_1,\ldots,I_k\},\, k>1}\sum_{j=1}^k:I_j:\, \prod_{l=1\atop l\not=j}^k\langle I_l\rangle^T_{\nu_0}.
\end{equation}
recursively defines\footnote{Note that  for $I=\{I_1,\ldots I_k\}$
with $k>1$, $\sharp I_j<\sharp X$ for $j=1,\ldots,k$.}  the Wick
ordered monomial $:X:=:\prod_{j\in X}\phi(x_j):_{\nu_0}$. }
\end{Def}

It remains to show that this definition  also solves the problem
of removing the self-con\-trac\-tions from the perturbation
series.  Let $J_1,\ldots J_m,X\subseteq \N$ be disjoint finite
sets. A partition $I\in {\cal P}(\cup_{l=1}^mJ_l\cup X)$,
$I=\{I_1,\ldots,I_k\}$, by definition has a self-contraction at a
set $J_l$, $l\in\{1,\ldots,m\}$, if $\exists j\in\N$, $1\leq j\leq k$, such that
$I_j\subseteq J_l$.  The collection of all partitions $I$ that do
not have self-contractions at $J_l$ for $l=1,\ldots,m$ is denoted
by ${\cal P}^{\rm Wick}(J_1,\ldots,J_m;X)$.
\begin{prop}
\label{5.1prop}
Let $J_1,\ldots J_m$ and $X$ as above. Then
\begin{equation}
\label{5.4eqa}
\langle :J_1:\cdots :J_m:\, X\rangle_{\nu_0}=\sum_{I\in{\cal P}^{\rm Wick}(J_1,\ldots,J_m;X)\atop I=\{I_1,\ldots I_k\}}\prod_{l=1}^k\langle I_l\rangle^T_{\nu_0}\, .
\end{equation}
\end{prop}
\noindent {\bf Proof.} The proof is by  induction over
$q=\sum_{l=1}^m\sharp J_l$. $q=0$ is just Definition \ref{3.1def}.
Suppose that (\ref{5.4eqa}) holds up to $q-1$. Then, by definition
of Wick ordering,
\begin{eqnarray}
\label{5.5eqa}
\langle :J_1:\cdots :J_m:\, X\rangle_{\nu_0}&=&\langle :J_1:\cdots :J_{m-1}: J_m\, X\rangle_{\nu_0}-\langle J_m\rangle_{\nu_0}\langle :J_1:\cdots :J_{m-1}:\, X\rangle_{\nu_0}\nonumber\\
&-&\sum_{Q\in{\cal P}(J_m)\atop Q=\{Q_1,\ldots, Q_k\},\, k>1}\sum_{j=1}^k\langle :J_1:\cdots :J_{m-1}::Q_j:X\rangle_{\nu_0}\prod_{l=1\atop l\not=j}^k\langle Q_l\rangle^T_{\nu_0}~.\nonumber\\
\end{eqnarray}
Application of the induction hypothesis  to the right hand side
yields
\begin{eqnarray}
\label{5.6eqa}
&&\sum_{I\in{\cal P}^{\rm Wick}(J_1,\ldots,J_{m-1};J_m\cup X)\atop I=\{I_1,\ldots I_k\}}\prod_{l=1}^k\langle I_l\rangle_{\nu_0}^T-\sum_{Q\in{\cal P}(J_m)\atop Q=\{Q_1,\ldots Q_k\}}\sum_{P\in{\cal P}^{\rm Wick}(J_1,\ldots,J_{m-1};X)\atop P=\{P_1,\ldots,P_{k'}\}}\prod_{l=1}^k\langle Q_l\rangle^T_{\nu_0}\prod_{l'=1}^{k'}\langle P_{l'}\rangle_{\nu_0}^T\nonumber\\
&&~~~~~~~~~~~~~~~~-\sum_{Q\in{\cal P}(J_m)\atop Q=\{Q_1,\ldots, Q_k\},\, k>1}\sum_{j=1}^k\sum_{P\in{\cal P}^{\rm Wick}( J_1,\ldots, J_{m-1},Q_j;X)\atop P=\{P_1,\ldots P_{k'}\}}\prod_{l'=1}^{k'}\langle P_{l'}\rangle_{\nu_0}^T\prod_{l=1\atop l\not=j}^k\langle Q_l\rangle^T_{\nu_0}~.
\end{eqnarray}
In the first sum we find all partitions  $I$ of $\cup_{l=1}^m
J_l\cup X$ that do not have self-contractions at $J_1,\ldots
J_{m-1}$. As in the second sum $I=Q\cup P$ is a partition of the
same set, we can identify this sum with the sum over all
partitions $I$ that do not have self-contractions at $J_1,\ldots,
J_{m-1}$ and where all points from $J_m$ are contained in
self-contractions. To complete the proof, the third sum finally
has to be identified with the sum over all partitions that do not
contain a self-contraction at $J_1,\ldots,J_{m-1}$ and do contain
at least one self-contraction at $J_m$, however not all points in
$J_m$ are being self-contacted.

Let $Q$, $j$ and $P$ be given from the  third sum. Then
$I=Q\setminus\{Q_j\}\cup P$ is such a partition: As
$Q\setminus\{Q_j\}\not=\emptyset$ there are self-contractions at
$J_m$, however the points in $Q_j\not=\emptyset$ are not contained
in a self-contraction.

Let, on the other hand, $I$ be a  partition of $ \cup_{l=1}^m
J_l\cup X$ from the set of partitions described above. Firstly,
for $Q\in{\cal P}(J_m)$, $\sharp Q=k$, we fix an enumeration
$1,\ldots,k$ of the elements of $Q$ (independently of $I$). Let
$\tilde Q=\{Q'\in I:Q'\subseteq J_m\}\not=\emptyset$, $
Q^c=J_m\setminus\cup_{Q'\in\tilde Q}Q'\not=\emptyset$ and
$k=\sharp \tilde Q+1>1$. Let $Q=\tilde Q\cup \{Q^c\}\in{\cal
P}(J_m)$ and $j$ be the number of the element $Q^c$. Furthermore,
we set $P=\{P'\cap [\cup_{l=1}^{m-1}J_l\cup Q^c\cup X]:P'\in
I\}\setminus\{\emptyset\}$. Then, $P\in {\cal P}^{\rm
Wick}(J_1,\ldots J_{m-1}, Q^c;X)$ and we get a map from the
prescribed set of partitions to the index set of the third sum.

It is easy to check that the two  maps between the described set
of partitions and the index set of the third sum of (\ref{5.6eqa})
(the other way round, respectively) that have been constructed in
the preceding two paragraphs are the inverses of each other. Hence
the correspondence between the two sets is one to one. Finally,
the contribution to the third sum determined by $Q,j$ and $P$
coincides with the contribution associated to the corresponding
$I=I(Q,j,P)$.  \prend

We can now define the $p$-th Wick  power
$:\phi^p:(x)=:\phi(x_1)\cdots\phi(x_n):|_{x_1,\ldots,x_n=x}$.
Obviously, $:\phi^p:(x)$ is a polynomial in the random variable
$\phi(x)$ with coefficients determined recursively according to
Def. \ref{5.1def} from the values of $C_n=\langle
\phi^n(x)\rangle_{\nu_0}^T$, $n<p$. By properties 1. and 2. of
$\nu_0$ (see Section \ref{2sec}), $C_n$ is finite and by property
3. it does not depend on $x$. Hence, $:\phi^p:=:\phi^p:_{\nu_0}$
is a well-defined polynomial in $\phi\in\R$. We also call this
polynomial the $p$-th Wick power. The main result of this section
is:

\begin{theo}
\label{5.1theo} Let $v(\phi)=\sum_{p=0}^{\bar p}\lambda_p\,
\phi^p$ and $:v(\phi):=\sum_{p=0}^{\bar p}\lambda_p :\phi^p:$. If
one replaces the interaction polynomial $v$ by its Wick-ordered
counterpart $:v:$, the perturbation series given in Theorem
\ref{3.1theo} remains the same with the only exception  that all
generalized Feynman graphs that contain self-contractions at inner
full vertices are removed from the series.
\end{theo}
\noindent {\bf Proof.} Note that a generalized Feynman graph has a
self-contraction at an inner full vertex if and only if the
corresponding partition (see Section \ref{3sec}) has a
self-contraction at the corresponding set of points $J_l$,
$l=1,\ldots,m$ (see also Figs.1 and 4). The theorem thus follows
from Proposition \ref{5.1prop}. \prend

For a centered Gaussian measure, Wick ordering of the  interaction
vertex means that no dashed line (see Fig. 3) leaving the vertex
can return to the same vertex, i.e. all Graphs containing a
sub-graph
\hspace{.3cm}\begin{picture}(0,0)\thicklines\put(0,.15){\circle*{.15}}\bezier{10}(0,.15)(.7,.45)(.8,.15)\bezier{10}(.8,.15)(.7,-.15)(0,.15)\put(1.05,0.03){$=$}\put(1.6,.15){\circle*{.15}}\put(2.55,.15){\circle{.16}}\bezier{100}(1.6,.15)(2.4,.45)(2.5,.25)\bezier{100}(2.5,.05)(2.4,-.15)(1.6,.15)\end{picture}\hspace{2.9cm}
are deleted from the perturbation series. This is of course the
well-known graphical meaning of Gaussian Wick-ordering.

In the centered Gaussian case,  Wick ordered monomials
$:J_1:=:\phi(x_1)\cdots\phi(x_n):$,
$:J_2:=:\phi(y_1)\cdots\phi(y_m):$ with a different number of
points $n\not=m$ are orthogonal in $L^2(\nu_0)$, as it is not
possible to make pairings out of
$\{x_1,\ldots,x_n,y_1,\ldots,y_m\}$ without getting at least one
self-contraction at $J_1=\{x_1,\ldots,x_n\}$ or
$J_2=\{y_1,\ldots,y_m\}$. If $n=m$, the only possible
contributions are those of Fig. 5 a), and hence $\langle
:J_1:\linebreak :J_2:\rangle_{\nu_0}=n!\,\langle{\rm
Sym}\otimes_{l=1}^n\delta_{x_l},{\rm
Sym}\otimes_{l=1}^n\delta_{y_l}\rangle_n$ with
$\langle.,.\rangle_n$ being the scalar product on ${\cal
H}^{\otimes n}$ and ${\cal H}$ the one particle Hilbert space
given by the closure of $C_0^{\infty}(\R^d)$ w.r.t. the inner
product $\langle u,h\rangle=\int_{\R^{2d}}u(x)h(y)\langle
\phi(x)\phi(y)\rangle_{\nu_0}\, dxdy$, $u,h\in C_0^\infty(\R^d)$.
Using property 1. from Section \ref{2sec}, it is easy to prove
that $\delta_x\in{\cal H}$. ${\rm Sym}$ stands for symmetrization.
As the span of $:J:$, $J\subseteq \N$ is dense in $L^2(\nu_0)$,
one obtains the Wiener-It\^o-Segal isomorphism between
$L^2(\nu_0)$ and the Bosonic Fock space over ${\cal H}$. For the
details we refer to \cite{GJ,Si}.

As a  self-contraction
\hspace{.3cm}\begin{picture}(0,0)\thicklines\put(0,.15){\circle*{.15}}\put(.05,.15){\line(1,0){.3}}\put(.4,.15){\circle{.15}}\end{picture}\hspace{.8cm}
can not occur at a Wick-ordered interaction vertex (or a
monomial), the above considerations also hold in the non-centered
Gaussian case where
\hspace{.3cm}\begin{picture}(0,0)\thicklines\put(0,0){$C_1=$}\put(1.05,.1){\line(1,0){.3}}\put(1.4,.1){\circle{.15}}\put(1.7,0){$\not=0$}\end{picture}\hspace{2.4cm}.

A functional  measure $\nu_0$ is non-Gaussian if and only if
$\exists n>2$ such that $\langle \phi(x_1)\cdots\linebreak
\phi(x_n)\rangle_{\nu_0}^T\not=0$ for some values of
$x_1,\ldots,x_n\in\R^d$. Let $m+1$ be the smallest such number and
$\{x_1,y_{1},\ldots,y_m\}$ be a collection of points such that the
truncated $m+1$-point function does not vanish. Obviously, the
$L^2(\nu_0)$ inner product of $:J_1:$ and $:J_2:$, $J_1=\{x_1\}$
and $J_2=\{y_1,\ldots,y_m\}$, consists out of only one non-zero
contribution depicted graphically in Fig. 5 b). For non-Gaussian
measures our graphical definition of Wick ordering does not give
an orthogonal decomposition of $L^2(\nu_0)$.

\begin{cor}
\label{5.1cor} Wick ordering  as defined in Def. \ref{5.1def}
gives an orthogonal decomposition of $L^2(\nu_0)$ in the sense of
a Wiener-It\^o-Segal isomorphism with the Bosonic Fock space if
and only if $\nu_0$ is Gaussian.
\end{cor}
\begin{figure}[t]
\begin{center}
\begin{picture}(12,2)
\thicklines
\put(-.2,.9){\bf a)}
\put(1,1){\oval(1,2)}
\put(3,1){\oval(1,2)}
\put(.8,1.4){$x_1$}
\put(.8,.4){$x_n$}
\put(.87,.8){$\vdots$}
\put(2.8,1.4){$y_{\pi_1}$}
\put(2.8,.4){$y_{\pi_n}$}
\put(2.85,.8){$\vdots$}
\put(1.5,1.5){\line(1,0){.44}}
\put(2,1.5){\circle{.15}}
\put(2.075,1.5){\line(1,0){.44}}
\put(1.5,.5){\line(1,0){.44}}
\put(2,.5){\circle{.15}}
\put(2.075,.5){\line(1,0){.44}}
\put(1.95,.8){\vdots}

\put(6.8,.9){\bf b)}
\put(8,1){\oval(1,2)}
\put(10,1){\oval(1,2)}
\put(7.85,.9){$x_1$}
\put(9.85,1.4){$y_1$}
\put(9.85,.4){$y_m$}
\put(9.92,.8){$\vdots$}
\put(8.5,1){\line(1,0){.44}}
\put(9,1){\circle{.15}}
\put(9.05,1.05){\line(1,1){.45}}
\put(9.05,.95){\line(1,-1){.45}}
\put(9.33,.8){$\vdots$}
\end{picture}
\end{center}
\caption{\footnotesize Contributions to  $\langle
:\!\phi(x_1)\cdots\phi(x_n)\!::\!\phi(y_1)\cdots\phi(y_m)\!\!:\,\rangle_{\nu_0}$
for a) the Gaussian case for $n=m$, $\pi\in{\rm Perm}(n)$, and b)
the non Gaussian case for $n=1,m>1$. }
\end{figure}
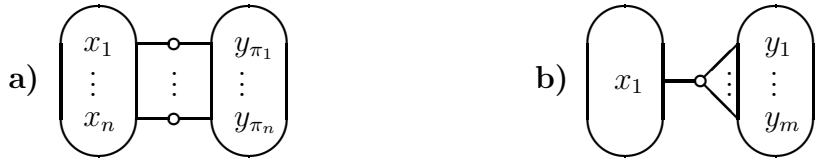

To close this section, let us consider  non-Gaussian some examples
in $d=2$ where Wick ordering renders the entire perturbation
series finite. If one however considers the non-Gaussian
generalization of Nelsons free field given by $L\phi=\eta$ with
$L=(-\Delta+m_0^2)^{1/2}$ and $\eta$ a non-Gaussian noise field,
i.e. $z>0$ in (\ref{4.1eqa}), see Section \ref{4sec}, one can
easily see from $g(x)\sim 1/|x|$ for small $|x|$ and Theorem
\ref{4.1theo} that Wick ordering does not remove all divergences:
Take, e.g. for a $:\phi^4:$-interaction in $d=2$ the generalized
Feynman graph in Fig. 1 which diverges logarithmically. For the
models of Section \ref{4sec}, Wick-ordering in $d=2$ thus is less
efficient than in the Gaussian case.

There is however a modification of these  models \cite{AGW3} where
Wick ordering in $d=2$ dimensions removes all divergences. Here we
briefly recall the construction. Let $\eta$ be a L\'evy noise
field, cf. Section \ref{4sec}, and let $\tilde\eta$ be a Gaussian
field with characteristic functional ${\cal
C}_{\tilde\eta}(h)=\exp\{-{c_2\over m_0^2}\int_{\R^d}|\nabla
h|^2\, dx\}$, $h\in{\cal S}$ with $c_2$ as in (\ref{4.3eqa}).
$\nabla$ is the gradient on $\R^d$. $\tilde \eta$ still is an infinite divisible, ultralocal field,
i.e. a field that has no correlations at a distance. We study the
linear SPDE $L\phi=\eta+\tilde\eta$ for $L=(-\Delta+m_0^2)$ where
$\eta$ and $\tilde \eta$ are assumed to be independent. Let $g$ be
the Greens function of $L$. Then the solution $\phi$ of this SPDE
has characteristic functional
\begin{equation}
\label{5.7eqa}
{\cal C}_{\nu_0}(h)=\exp\{\int_{\R^d} [\psi(g*h)-{c_2\over m^2_0}|\nabla (g*h)|^2]\, dx\}~,~h\in{\cal S}.
\end{equation}
Performing the functional derivatives of  $\log {\cal
C}_{\nu_0}(h)$ at $h=0$ one obtains that the truncated moment
functions for $n\not=2$ are given by (11). For $n=2$ one obtains
due to the correction term induced by $\tilde\eta$:
$\langle\phi(x)\phi(y)\rangle^T_{\nu_0}=\tilde c_2\,g(x-y)$ with
$\tilde c_2=c_2/m_0^2$.

The Feynman rules of Theorem \ref{4.1theo}  now change as follows:
Put a propagator $\tilde c_2 g(y-y')$ for all subgraphs
$y~\mbox{\bf -}\hspace{-.15cm}\circ\hspace{-.13cm}\mbox{\bf
-}~y'$. Then proceed as in Theorem \ref{4.1theo} for the remaining
vertices and edges\footnote{An alternative description is to draw
generalized Feynman graphs with two kinds of edges, dotted and
not, and inner empty vertices that have three or more legs (for
$\eta$ centered). Dotted edges go from full to full vertices,
non-dotted edges from empty to full vertices. Then the graph can
be evaluated as in Theorem \ref{4.1theo} if one multiplies with an
extra $\tilde c_2$ for each dotted edge.}.  For the given $L$, the
singularity of $g(x)$ at $x=0$ is only logarithmic $\sim -{1\over
2\pi}\log|x|$ and $g(x)\sim e^{-m_0|x|}$ for $|x|$ large. As there
are no self-contractions in a Wick ordered gen. Feynman graph and
arbitrary powers of $g$ are integrable, one obtains:

\begin{theo}
\label{5.2theo} Let $d=2$ be the dimension of the  underlying
space, $\nu_0$ constructed as above and
$:v(\phi):=\sum_{p=0}^{\bar p}\lambda_p:\phi^p:$ the interaction
density. Then the perturbation series of $\nu_0$ is free of
divergences, i.e. the perturbation series of the ultra-violet
regularized measures $\nu_0^\epsilon$ with interaction densities
$:v(\phi):_\epsilon=\sum_{p=0}^{\bar p}\lambda_p
:\phi^p:_{\nu_0^\epsilon}$ converges term by term as
$\epsilon\searrow 0$.
\end{theo}
\noindent {\bf Proof.} Let $G$ be a generalized  Feynman graph. By
the Feynman rules described above, all the values ${\cal
V}_\Lambda[G](x_1,\ldots,x_n)$ (up to constants) occur also in the
perturbation theory of some Gaussian $P(\phi)_2$-theory. The proof
thus is essentially\footnote{Here, infra-red cut-offs have to be
treated slightly more carefully as empty vertices do not have such
a cut-off. One can take this into account by integrating first
over the empty vertices and then over the full ones. Note that every connected component
of a generalized Feynman graph contains at least one full vertex that provides an IR-cut-off.} the same as
in \cite{GJ} Lemma 8.5.2 and Theorem 8.5.3. \prend

We note that the models described in Theorem  \ref{5.2theo} in
particular include Nelson's free field, take $z=0$ in
(\ref{4.1eqa}). For $z>0$, the measures $\nu_0$ are non-Gaussian.
Even though the truncated moment functions
$\langle\phi(x_1)\cdots\phi(x_n)\rangle^T_{\nu_0}$ of such $\nu_0$
are continuous functions for $n\geq 3$ and hence the constants
$C^\epsilon_n=\langle\phi^n(x)\rangle^T_{\nu_0^\epsilon}$ are
finite in the limit $\epsilon \searrow 0$, one cannot replace the
Wick ordering in the perturbation series w.r.t. $\nu_0$ with the
Wick-ordering w.r.t. a Gaussian measure with the same covariance
functions if one wants to get a finite perturbation series. We
take e.g. a $:\phi^p:_{\nu_0^\epsilon}$ interaction for a
symmetric measure (cf. Corollary \ref{3.2cor}). For $p=2$ Gaussian
and non-Gaussian Wick ordering coincide. For $p=4$ they still
coincide up to a constant $C_4^\epsilon$ that converges for
$\epsilon\searrow 0$ and can be neglected. For $p=6$ however, in
the difference there is an additional constant term $\sim
C_4^\epsilon C_2^\epsilon$ that diverges, but this can still be
considered as an irrelevant ground state energy. Finally, for
$p=8$ there is a logarithmically divergent mass-counterterm $\sim
C_4^\epsilon C_2^\epsilon :\phi^2:_{\nu_0^\epsilon}$ present in
the non-Gaussian Wick ordering that is missing in the Gaussian
one. This makes it clear that one cannot hope for a finite
perturbation series using the wrong (Gaussian) Wick ordering if
$p\geq 8$.

\section{Linked cluster theorem for generalized Feynman graphs}
\label{6sec} In this section we solve the Problem 2 and the  first
part of Problem 4 of Section \ref{2sec}, i.e. we perturbatively
calculate the free energy density $f_\Lambda=\log
Z_\Lambda/|\Lambda|$ and we prove the existence of the
thermodynamic (TD) limit $\Lambda\nearrow\R^d$ for each term in
the perturbation series for a general $\nu_0$ with a sufficiently
fast clustering, cf. property 4 of Section \ref{2sec}, F2. and Eq.
(\ref{3.2eqa}). The result is the expected one -- only connected
generalized Feynman graphs contribute to $f_\Lambda$ -- and can be
seen as one of the many variations of the linked cluster theorem.
As the method of proof, we do not use polymer systems, see e.g.
\cite{Ba,Sa}, but use bookkeeping of partitions instead.

First we note that  $Z_\Lambda(\Lambda,\beta\lambda_0,\ldots\beta\lambda_{\bar p} )=\langle e^{-\beta
V_\Lambda}\rangle_{\nu_0}$ is the Laplace transform of the random
variable $V_\Lambda$ in the parameter $\beta>0$. If we want to
expand into powers $V_\Lambda$, we can expand in powers of $\beta$
and put $\beta=1$ afterwards. The this expansion of course is the
one obtained in Section \ref{3sec} for $n=0$. If we now want to
expand the free energy density
$f_\Lambda=f(\Lambda,\lambda_0,\ldots,\lambda_{\bar p})=\log
Z_\Lambda(\Lambda,\lambda_0,\ldots,\lambda_{\bar p})/|\Lambda|$ into powers of $V_\Lambda$, we can
do the same for $f(\Lambda,\beta\lambda_0,\ldots,\beta\lambda_{\bar p})$. By the basic linked
cluster theorem F4 in Section \ref{4sec}, see also Appendix A, we
get in the sense of formal power series
\begin{equation}
\label{6.1eqa}
\log Z_\Lambda=\sum_{m=1}^\infty {(-1)^m\over m!}\, \langle V_\Lambda^m\rangle_{\nu_0}^{(T)}\, ,
\end{equation}
where the superscript  $(T)$ means that in the combinatorics of
Def. \ref{3.1def} each of the $m$ copies of $V_\Lambda$ is treated
as one object, even if they contain higher powers of the field
variables $\phi(x)$. As already in the preceding section, the
superscript $T$ is reserved for the combinatorics in Def.
\ref{3.1def} where each copy of $\phi(x)$ is treated as one
object. As the latter combinatorics is linked with generalized
Feynman graphs, we have to expand $\langle
V_\Lambda\rangle^{(T)}_{\nu_0}$ in terms of truncated moments
$\langle \phi(x_1)\cdots\phi(x_n)\rangle^T_{\nu_0}$.  The first
step is to prove the $(T)$-truncated analogue of Eq.
(\ref{2.3eqa}), i.e. that one can interchange the truncated
expectation and the integrals over $\Lambda$:

\begin{lem}
\label{6.1lem} For $V_\Lambda$ defined as in Eq. (\ref{2.1eqa}) the following holds:
\begin{equation}
\label{6.2eqa}
\langle V_\Lambda^m\rangle^{(T)}_{\nu_0}=\sum_{p_1,\ldots,p_m}^{\bar p}\int_{\Lambda^m}\lambda_{p_1}\cdots\lambda_{p_m}\langle\phi^{p_1}(y_1)\cdots\phi^{p_m}(y_m)\rangle^{(T)}_{\nu_0}\, dy_1\cdots dy_m\, .
\end{equation}
Here the superscript $(T)$  means that in Def. \ref{3.1def} each
random variable $\phi^{p_l}(y_l)$, $l=1,\ldots,m$, and each of the
$m$ copies of $V_\Lambda$ is being treated as one object.
\end{lem}
\noindent {\bf Proof.} By  Fubini's theorem (\ref{6.2eqa}) is true
if we omit the $(T)$ on both sides. For $m=1$ the $\langle
V_\Lambda\rangle^{(T)}_{\nu_0}=\langle V_\Lambda\rangle_{\nu_0}$
and $\langle
\phi^{p_1}(y_1)\rangle^{(T)}_{\nu_0}=\langle\phi^{p_1}(y_1)\rangle_{\nu_0}$.
Hence (\ref{6.2eqa}) also holds for $m=1$. For $m>1$ we get by
induction and Def. \ref{3.1def} that the difference between the
left hand side and the right hand side of (\ref{6.2eqa}) without
$(T)$ consists only out of the truncated terms on both sides with
the partition $I=\{\{1,\ldots,m\}\}$ and hence out of the
difference of both sides of (\ref{6.2eqa}) with the superscript
$(T)$. This difference must thus be zero. \prend

Let $J_1,\ldots,J_m\subseteq \N$  be disjoint sets. By definition,
a partition $I\in{\cal P}(\cup_{l=1}^mJ_l)$ is connected w.r.t.
the "blocks" $J_1,\ldots ,J_m$, in notation $I\in{\cal
P}_c(J_1,\ldots,J_m)$, if for $I=\{I_1,\ldots,I_k\}$ $\not
\!\!\exists \, 1\leq i_1,\ldots,i_q\leq k$, $1\leq q<k$ and $1\leq
j_1,\ldots,j_s\leq m$, $1\leq s< m$ such that $\cup_{\alpha=1}^q
I_{i_\alpha}=\cup_{\alpha=1}^sJ_{j_\alpha}$. Let $X\subseteq\N$ be
the set outer full vertices, $X\cap J_l=\emptyset$,
$l=1,\ldots,m$. A partition $I\in{\cal P}(\cup_{l=1}^m J_l\cup X)$
is connected w.r.t the blocks $J_1,\ldots,J_m$ and the outer
points $X=\{k_1,\ldots,k_n\}$ if $I\in{\cal
P}_c(J_1,\ldots,J_m,\{k_1\},\ldots,\{k_n\})$. We then write $I\in
{\cal P}_c(J_1,\ldots,J_m;X)$.

A graph $G$ is connected, if there  exists an enumeration of its
vertices such that each two subsequent vertices are connected by
an edge. The set of connected generalized Feynman graphs with $n$
outer full vertices and $m$ inner full vertices is denoted by
${\cal F}_c(n,m)$.

\begin{lem}
\label{6.2lem} A generalized Feynman graph  $G\in{\cal F}(n,m)$ is
connected if and only if the partition associated to $G$, cf.
Lemma \ref{3.1lem} and Fig. 1, is connected w.r.t. the blocks
$J_1,\ldots,J_m$ of the legs of the inner full vertices and points
$X$ of the outer full vertices.
\end{lem}
\noindent {\bf Proof.} As we can treat the  points in
$X=\{k_1,\ldots,k_n\}$ as $n$ additional blocks
$J_{m+1}=\{k_1\},\ldots,J_{m+n}=\{k_n\}$, it suffices to prove the
statement for $n=0$.

Let $G\in{\cal F}_c(0,m)$ and $I=\{I_1,\ldots,I_k\}$ be the
associated partition in ${\cal P}(\cup_{l=1}^mJ_l)$. From Section
\ref{3sec} it is clear that there exists a bijection between the
full vertices of $G$ and the sets $\{J_1,\ldots J_m\}$ and between
the empty vertices and the sets $\{I_1,\ldots,I_k\}$. Let
$\{i_1,\ldots i_q\}\subseteq \{1,\ldots,k\}$ and
$\{j_1,\ldots,j_s\}\subseteq \{1,\ldots,m\}$ such that
$\cup_{\alpha=1}^qI_{i_\alpha}=\cup_{\alpha=1}^sJ_{j_\alpha}$.
Then all edges from inner full vertices associated to one $J_j$
with index $j$ in $\{j_1,\ldots, j_s\}$ go to an empty vertex
associated with an $I_i$ with $i\in\{i_1,\ldots,i_q\}$ and vice
versa. Hence no edge leaves/comes into the subgraph $G'\subseteq
G$ that consists out of the full vertices labelled by
$\{j_1,\ldots j_s\}$ and the empty ones labelled by $\{i_1,\ldots
i_q\}$ and all the edges between these vertices. By connectedness
of $G$, $G'=G$. Hence $\{i_1,\ldots, i_q\}=\{1,\ldots ,k\}$ and
$\{j_1,\ldots,j_s\}=\{1,\ldots,m\}$.

Conversely, let $I\in{\cal P}_c(J_1,\ldots,J_m)$,
$I=\{I_1,\ldots,I_k\}$, be connected and $G$ the associated
generalized Feynman graph and $G'$ be a maximal connected subgraph
of $G$. Let $\{i_1,\ldots i_q\}$ and $\{j_1,\ldots,j_s\}$ be the
index sets of the empty respectively full vertices of $G'$
obtained through the identification of full inner vertices with
sets $\{J_1,\ldots,J_m\}$ and empty inner vertices with
$\{I_1,\ldots I_k\}$. As $G'$ is maximal, all edges in $G$
connected to a vertex in $G'$ are in $G'$, hence
$\cup_{\alpha=1}^qI_{i_\alpha}=\cup_{\alpha=1}^sJ_{j_\alpha}$.
From the connectedness of $I$ it follows that $\{i_1,\ldots,i_q\}=
\{1,\ldots,k\}$ and $\{j_1,\ldots,j_s\}=\{1,\ldots,m\}$. Hence
$G'=G$. \prend

\begin{prop}
\label{6.1prop} Let $J_1,\ldots,J_m\subseteq\N$ be disjoint sets
and $\langle J_1\cdots J_m\rangle^{(T)}_{\nu_0}$ be the truncated
moment where each random variable $J_1,\ldots,J_m$ in the
combinatorics of Def \ref{3.1def} is treated as one object. Then
this "block truncated" moment has the following expansion into
truncated moments
$\langle\phi(x_1)\cdots\phi(x_n)\rangle^T_{\nu_0}$:
\begin{equation}
\label{6.3eqa}
\langle J_1\cdots J_m\rangle^{(T)}_{\nu_0}=\sum_{I\in{\cal P}_c(J_1,\ldots,J_m)\atop I=\{I_1,\ldots,I_k\}}\prod_{l=1}^k\langle I_l\rangle^T_{\nu_0}\,.
\end{equation}
\end{prop}
\noindent {\bf Proof.} Note that the ordinary moment functions
determine the (block) truncated moments and vice versa. Hence,
(\ref{6.3eqa}) holds if and only if the right hand side of this
equation fulfills the defining equation for the left hand side,
i.e. if and only if for all $m\in\N$
\begin{equation}
\label{6.4eqa}
\langle J_1\cdots J_m\rangle_{\nu_0}=\sum_{I\in {\cal P}\{1,\ldots,m\}\atop I=\{I_1,\ldots,I_k\}}\prod_{l=1}^k\left[ \sum_{Q_l\in{\cal P}_c(J_q:q\in I_l)\atop Q_l=\{Q_{l,1},\ldots Q_{l,k_l}\}}\prod_{s_l=1}^{k_l}\langle Q_{l,s_l}\rangle^T_{\nu_0}\right]
\end{equation}
holds. Given $\tilde I=\{i_1,\ldots,i_s\}\subseteq
\{1,\ldots,m\}$,  we have introduced the notation ${\cal
P}_c(J_q:q\in \tilde I)$ for ${\cal P}_c(J_{i_{1}},\ldots,
J_{i_{s}})$.

On the other hand,  we can expand the left hand side of
(\ref{6.4eqa}) into truncated moment functions
\begin{equation}
\label{6.5eqa}
\langle J_1\cdots J_m\rangle_{\nu_0}=\sum_{R\in{\cal P}(\cup_{l=1}^mJ_l)\atop R=\{R_1,\ldots,R_k\}}\prod_{l=1}^k\langle R_l\rangle^T_{\nu_0}
\end{equation}
and we have to prove  that the right hand side of (\ref{6.4eqa})
equals the right hand side of (\ref{6.5eqa}).

Given $I\in{\cal P}\{1,\ldots,m\}$, $I=\{I_1,\ldots,I_k\}$ and
$Q_l\in{\cal P}_c(J_q:q\in I_l)$ for $l=1,\ldots,k$ one gets a
partition $R=R(I,Q_1,\ldots,Q_k)$ from ${\cal P}(\cup_{l=1}^mJ_l)$
setting $R=\cup_{l=1}^kQ_l$.  The corresponding contributions to
the right hand side of (\ref{6.4eqa}) and (\ref{6.5eqa}) are
obviously equal. It remains to prove that the mapping
$R(I,Q_1,\ldots,Q_k)$ from the index set of the total sum on the
right hand side of (\ref{6.4eqa}) to ${\cal P }(\cup_{l=1}^mJ_l)$
is one to one.

Again, this can be proven by  construction of the inverse mapping.
Let $R\in {\cal P}(\cup_{l=1}^mJ_l)$ be given. For $1\leq q<j\leq
m$ we say that $R$ connects $q$ and $j$, in notation $q\sim_Rj$,
if the full inner vertices corresponding to $J_q$ and $J_j$,
respectively, are connected in the generalized Feynman graph
corresponding to $R$, cf. Section \ref{3sec}. Obviously, $\sim_R$
is an equivalence relation on $\{1,\ldots,m\}$. Let
$I=\{I_1,\ldots,I_k\}$ be the equivalence classes of $\sim_R$,
then $I\in{\cal P}\{1,\ldots,m\}$. For $l=1,\ldots,k$, let
$Q_l=\{\tilde R\in R:\tilde R\subseteq \cup_{q\in I_l}J_q\}$. It
remains to show that $Q_l\in{\cal P}_c(J_q:q\in I_l)$.

Firstly,  $Q_l=\{Q_{l,1},\ldots,Q_{l,k_l}\}\in{\cal P}(\cup_{q\in
I_l}J_q)$. If not, then there are some points in $\cup_{q\in
I_l}J_q$ that are not in $\cup_{s_l=1}^{k_l} Q_{l,s_l} $. A set
$\tilde R\in R$ that contains at least one of these points, say
from $J_q$ for $q\in I_l$, can not contain any point from $J_j$,
$j\not\in I_l$, as this would imply that one can go in the graph
corresponding to $R$ from the full inner vertex $J_q$ to $J_j$ via
the empty vertex $\tilde R$ in contradiction with $q\not\sim_R j$.
Hence $\tilde R\in Q_l$, but this contradicts the assumption that
$\tilde R$ contains at least one  element $\not \in
\cup_{s_l=1}^{k_l}Q_{l,s_l}$.

 Secondly,  $Q_l$ is a connected partition
with respect  to $J_q, q\in I_l$, as the subgraph with full inner
vertices $J_q$, $q\in I_l$, and empty inner vertices $Q_{l,s_l}$,
$s_l=1,\ldots,k_l$, in the graph associated to $R$ by definition
of $\sim_R$ is connected. An application of Lemma \ref{6.2lem}
therefore concludes the proof. \prend

Combination of  Lemmas \ref{6.1lem}, \ref{6.2lem} and Proposition
\ref{6.1prop} now gives the general linked cluster theorem:

\begin{theo}
\label{6.1theo} The perturbations  series of \, $\log Z_\Lambda$
for the energy density $v(\phi)=\linebreak \sum_{p=0}^{\bar
p}\lambda_p\, \phi^p$ only contains the connected generalized
Feynman graphs, i.e. in the sense of formal power series one gets
\begin{equation}
\label{6.6eqa}
\log Z_\Lambda=\sum_{m=1}^\infty {(-1)^m\over m!}\sum_{G\in {\cal F}_c(0,m)}{\cal V}_\Lambda[G]\,.
\end{equation}
\end{theo}

Let ${\cal P}_c^{\rm Wick}(J_1,\ldots,J_m;X)$  be the intersection
of ${\cal P}_c(J_1,\ldots,J_m;X)$ and ${\cal P}^{\rm
Wick}(J_1,\ldots,\linebreak J_m;X)$ and let ${\cal F}^{\rm
Wick}_c(n,m)$ be the collection of connected generalized Feynman
graphs without self-contractions at the inner full vertices. The
generalization of Lemmas \ref{6.1lem}, \ref{6.2lem} and Prop.
\ref{6.1prop} to the Wick ordered case is straight forward. One
obtains the Wick ordered version of the general linked cluster
theorem:

\begin{cor}
\label{6.1cor} If one replaces $v(\phi)$ by its Wick ordered
counterpart, (\ref{6.6eqa}) still holds if one restricts the sum
on the right hand side to ${\cal F}_c^{\rm Wick}(0,m)$.
\end{cor}

The main  application of linked cluster expansions in statistical
mechanics is to prove the existence of the free energy density in
the TD limit and to obtain an approximative  formula for it. Here,
for simplicity, we restrict to short range forces. The adequate
formulation is as follows: Let $\nu_0$ be a measure with
exponential clustering, i.e. $\exists m_0>0$ such that for
$X,Y\subseteq \N$, $|\langle XY\rangle_{\nu_0}-\langle
X\rangle_{\nu_0}\langle Y\rangle_{\nu_0}|\leq
D\exp\{-m_0\,\underline{d}(X,Y)\}$ where $D$ is a constant
depending only\footnote{In the non uv-regular situation things are
getting slightly more complicated, cf. Section \ref{8sec}.} on
$\sharp X$ and $\sharp Y$ and ${\underline
d}(X,Y)=\min\{|x_j-y_l|:j\in X,l\in Y\}$ is the minimal distance
between the points in $X$ and $Y$. It is well-known, see e.g.
\cite{Ru}, that this is equivalent with $|\langle X
Y\rangle^T_{\nu_0}|\leq D'\exp\{-m_0\,\underline{d}(X,Y)\}$ for
$D'=D'(\sharp X,\sharp Y)$ another constant. This is just a more
precise statement of F2. For the convenience of the reader we give
a proof of this statement in Appendix A.

The TD limit  is to let $\Lambda\nearrow\R^d$ in the sense of Van Hove, cf \cite[p.14]{Ru} and Appendix \ref{Bsec} below.

\begin{theo}
\label{6.2theo} Let $\nu_0$ be  a measure with exponential
clustering and $v(\phi)=\sum_{p=0}^{\bar p}\lambda_p\,\phi^p$ the
energy density. Then the perturbation series for the free energy
density $f_\Lambda$ converges in the sense of formal power series.
The limit $f=\lim_{\Lambda\nearrow \R^d}f_\Lambda$ is given by
\begin{equation}
\label{6.7eqa}
f=\sum_{m=1}^\infty{(-1)^m\over m!}\sum _{G\in {\cal F}_c(0,m)}{\cal V}'[G]
\end{equation}
where ${\cal V}'[G]$ is obtained  from the same Feynman rules as
${\cal V}_\Lambda$, cf. Theorem \ref{3.1theo}, with the only
difference that the integration over one inner full vertex is
omitted\footnote{Note that by the translation invariance of
$\nu_0$, the result for ${\cal V}'$ does not depend on the
argument $y_l\in\R^d$ or the choice $l=1,\ldots,m$ of this inner
full vertex.} and $\Lambda$ in the remaining integrations is
replaced by $\R^d$.

If one Wick orders $v(\phi)$, (\ref{6.7eqa})  still holds for the
sum on the right hand side restricted to graphs without
self-contractions.
\end{theo}
\noindent {\bf Proof.} We have to prove that  $\lim_{\Lambda
\nearrow \R^d}{\cal V}_\Lambda[G]/|\Lambda|={\cal V}'[G]$ for all
$G\in{\cal F}_c(0,m)$. We thus have to  prove that the integrand
in the Feynman rules for $G$ fulfills the conditions on
$I(y_1,\ldots,y_m)$ in Appendix B. Obviously, it is translation
invariant. As the measures under discussion are uv-regularized,
one can prove (\ref{B.0eqa}) for $B=0$.

Let $y_1$ and $y_2$ be the values attached to  two inner full
vertices. As $G$ is connected, there is a path on $G$ from $y_1$
to $y_2$ passing through at most $0<m_1<m-2$ inner full vertices
and $0<m_2<m-1$ inner empty vertices. On the path from $y_1$ to
$y_2$ there must be at least one of the $m_1$ steps from one inner
full vertex to its successor vertex of the same kind that is $\geq
|y_1-y_2|/m_1$. Let $n$ be the number of legs of the inner empty
vertex that has been passed during this step. Then the $n$
arguments of the corresponding truncated function can be divided
into two groups with a minimal mutual distance of
$|y_1-y_2|/m_1n$. By the cluster hypothesis this leads to a decay
of the integrand $\leq C\exp\{-(m_0/m_1n)|y_1-y_2|\}$ for $C$
sufficiently large.

Let now $y_1=0$ and $y_2,\ldots,y_m$ be the remaining values
assigned to the inner full vertices. Note that
$\sum_{l=2}^m|y_l|\leq (m-1)\max\{|y_l|:L=2,\ldots,m\}$, hence the
integrand of ${\cal V}_\Lambda(G)$ fulfills the estimate
(\ref{B.0eqa}) for $M=m_0/(m^2\bar n)$ where $\bar n$ is the
maximal number of legs at an empty vertex in ${\cal F}(0,m)$, i.e.
$\bar n=m\bar p$. \prend

It is easy to verify the exponential clustering for the models of
Section 4 provided that $|g(x)|\leq D'' \exp\{-m_0|x|\}$, see (11)
and also \cite{AGW1}. Hence Theorem \ref{6.1theo} applies to these
measures. This is also true for the two-dimensional models
described in Theorem \ref{5.2theo}, as the proof of the above
theorem can be easily adapted to the case where logarithmic
divergences occur at coinciding points of the integrand, cf.
Appendix \ref{Bsec}.

\section{TD limit of (truncated) moment functions}
\label{7sec} In this short  section we apply the results of
Section \ref{6sec} to the generating functionals of the truncated
moment functions of the interacting measure $\nu_\Lambda$ in order
to complete the solution of Problems 3 and 4 of Section
\ref{2sec}. Apart from the input from Section \ref{6sec}, the
methods we use here are more or less standard, see e.g. \cite{Ba}.

For $h\in {\cal S}$, let $v_h(\phi)=\sum_{p=0}^{\bar
p}(\lambda_p-i\delta_{1,p}h)\phi^p$ be the energy density
$v(\phi)=\sum_{p=0}^{\bar p}\lambda_p\,\phi^p$ with an additional
"Schwinger term\footnote{Here the imaginary unit $i=\sqrt{-1}$ in
front of the Schwinger term has been chosen in order to match with
our conventions that the generating functional is the
characteristic function, i.e. the (functional) Fourier transform
and not the Laplace transform.}" $-ih\phi$.  Let
$V_{\Lambda,h}(\phi)=\int_{\Lambda}v_h(\phi)\,dx$ and
$Z_\Lambda(h)=\langle e^{-V_{\Lambda,h}}\rangle_{\nu_0}$.
Obviously, ${\cal C}_{\nu_\Lambda}(h)=Z_\Lambda^{-1}\langle
e^{i\langle
\phi,h\rangle}e^{-V_\Lambda}\rangle_{\nu_0}=Z_\Lambda(h)/Z_\Lambda$
for $h\in{\cal S}$, ${\rm supp}h\subseteq \Lambda$. Hence, ${\cal
C}^T_{\nu_\Lambda}(h)=\log Z_\Lambda(h)-\log Z_\Lambda$ and
\begin{equation}
\label{7.1eqa}
\langle\phi(x_1)\cdots\phi(x_n) \rangle_{\nu_\Lambda}^T=(-i)^n\left.{\delta^n \log Z_\Lambda(h)\over\delta\phi(x_1)\cdots\delta\phi(x_n)}\right|_{h=0}~~{\rm for}~x_1,\ldots,x_n\in\Lambda,~n\in\N.
\end{equation}
We  want to find a graphical expression for (\ref{7.1eqa}). Let
${\cal F}^{\rm Sw.}_c(m)$ be the collection of connected
generalized Feynman graphs without outer vertices and with one
additional type of inner full vertex (henceforth called Schwinger
vertex) such that the total number of inner full vertices is $m$.
The additional vertex type has one leg and corresponds to the
Schwinger term. For $G\in{\cal F}^{\rm Sw.}_c(m)$ and $h\in{\cal
S}$ let ${\cal V}_\Lambda[G](h)$ be the value obtained according
to the Feynman rules Def. \ref{3.3def} where $-ih$ is the coupling
constant for the additional one-legged vertex. Then, by Theorem
\ref{6.1theo}, $\log Z_\Lambda(h)=\sum_{m=1}^\infty{(-1)^m\over
m!}\sum_{G\in{\cal F}^{\rm Sw.}_c(m)}{\cal V}_\Lambda[G](h)$ holds
in the sense of power series in the formal parameters
$\lambda_0,\ldots,\lambda_{\bar p},h$. Inserting this into
(\ref{7.1eqa}), one gets for $x_1,\ldots,x_n\in\Lambda$, $n\in\N$
\begin{equation}
\label{7.2eqa}
\langle\phi(x_1)\cdots\phi(x_n) \rangle_{\nu_\Lambda}^T=(-i)^n\sum_{m=1}^\infty{(-1)^m\over m!}\sum_{G\in{\cal F}^{\rm Sw.}_c(m)}\left.{\delta^n {\cal V}_\Lambda[G](h)\over\delta\phi(x_1)\cdots\delta\phi(x_n)}\right|_{h=0}.
\end{equation}
For $G\in {\cal F}_c^{\rm Sw.}(m)$ let $n'$ be the number of  the
one-legged inner full vertices corresponding to the Schwinger term
in the energy density. Then, $\left.{\delta^n {\cal
V}_\Lambda[G](h)\over\delta\phi(x_1)\cdots\delta\phi(x_n)}\right|_{h=0}=0$
if $n'\not=n$ and
\begin{equation}
\label{7.3eqa}
\left.{\delta^n {\cal V}_\Lambda[G](h)\over\delta\phi(x_1)\cdots\delta\phi(x_n)}\right|_{h=0}=(-i)^n\sum_{\sigma\in{\rm Perm}(n)}{\cal V}_\Lambda[G'](x_{\sigma_1},\ldots,x_{\sigma_n})
\end{equation}
if $n'=n$ and $G'\in {\cal F}_c(n,m-n)$ is the graph obtained
from $G$ by replacing all Schwinger vertices with outer full
vertices. ${\rm Perm}(n)$ is the permutation group of $n$ objects.
Obviously, all graphs from ${\cal F}_c(n,m-n)$ can be obtained in
this way from some $G\in {\cal F}_c^{\rm Sw.}(m)$.

For a generic $G\in{\cal F}_c^{\rm Sw.}(m)$, each of its $m$
inner full vertices can be a Schwinger vertex or not. If one has
to choose exactly $n$ from $m$ vertices to become Schwinger
vertices, there are thus $({m\atop n})$ possibilities. Once this
choice has been done, every permutation of the $n$ Schwinger
vertices and the $m-n$ remaining inner full vertices leads to a
distinct $G'$ as full vertices are distinguishable. The covering
$\{ G\in{\cal F}^{\rm Sw.}_c: G$ has $n$ Schwinger vertices$\}\ni
G\to G'\in {\cal F}_c(n,m-n)$ thus is $({m\atop n})$-fold. Using
this and inserting (\ref{7.3eqa}) into (\ref{7.2eqa}) one obtains
\begin{equation}
\label{7.4eqa}
\langle\phi(x_1)\cdots\phi(x_n)\rangle^T_{\nu_\Lambda}=(-i)^{2n}\sum_{m=n}^\infty{(-1)^m\over m!}\,n!\left({m\atop n}\right)\sum_{G\in{\cal F}_c(n,m-n)}{\cal V}_{\Lambda}[G](x_1,\ldots,x_n).
\end{equation}
The factor $n!$ stems from the sum over ${\rm Perm} (n)$ in
(\ref{7.3eqa}). Re-arranging (\ref{7.4eqa}) in powers of the
formal parameters $\lambda_0,\ldots,\lambda_{\bar p}$ then yields
\begin{theo}
\label{7.1theo} For $n\in\N$ and $x_1,\ldots,x_n\in\Lambda$ the
truncated moment functions of the interacting measure
$\nu_\Lambda$  are given by the formal power series
\begin{equation}
\label{7.5eqa}
\langle\phi(x_1)\cdots\phi(x_n)\rangle^T_{\nu_\Lambda}=\sum_{m=0}^{\infty}{(-1)^m\over m!}\sum_{G\in {\cal F}_c(n,m)}{\cal V}_\Lambda[G](x_1,\ldots,x_n).
\end{equation}
Let  $\nu_0$ have the exponential clustering property. Then,  the
right hand side of (\ref{7.5eqa}) converges to
$\langle\phi(x_1)\cdots\phi(x_n)\rangle_\nu^T$ as
$\Lambda\nearrow\R^d$ which is understood here as the formal power
series given on the r.h.s. of (\ref{7.5eqa}) with ${\cal
V}_\Lambda[G](x_1,\ldots,x_n)$ replaced by ${\cal
V}[G](x_1,\ldots,x_n)$. The latter expression is obtained through
the Feynman rules as in Definition \ref{3.3def} with the
integration over the inner full vertices extended over all $\R^d$.
\end{theo}
\noindent {\bf Proof.} Only  the convergence in the TD limit
${\cal V}_\Lambda[G](x_1,\ldots,x_n)\to{\cal
V}[G](x_1,\ldots,x_n)$ as $\Lambda\nearrow\R^d$ needs to be
proven.

As $G$ is connected and $\nu_0$  is clustering exponentially fast,
one can apply arguments similar to those in the proof of Theorem
\ref{6.2theo} to prove that the integrand in ${\cal
V}_\Lambda[G](x_1,\ldots,x_n)$is of exponential decay if any of
the values attached to the inner full vertices becomes separated
from any of the outer points $x_1,\ldots,x_n\in\R^d$. Thus, the
assertion of the theorem follows from Lebesgue's theorem of
dominated convergence. \prend

Let $\tilde {\cal F}(n,m)$ be the collection of generalized
Feynman graphs with $m$ inner full vertices and $n$ outer full
vertices such that any connected component of $G$ contains at
least one outer full vertex. Using Theorem \ref{7.1theo} in
combination with Def. \ref{3.1def} gives:
\begin{cor}
\label{7.1cor} As a formal power series, the moment functions of
the interacting  measure $\nu=\lim_{\Lambda \nearrow
\R^d}\nu_\Lambda$ are given by
\begin{equation}
\label{7.6eqa}
\langle\phi(x_1)\cdots\phi(x_n)\rangle_\nu=\sum_{m=0}^\infty {(-1)^m\over m!}\sum_{G\in\tilde{\cal F}(n,m)}{\cal V}[G](x_1,\ldots,x_n).
\end{equation}
\end{cor}
This completes the solution of Problem 4 in Section \ref{2sec}.
Clearly, when replacing $v(\phi)$ with its Wick-ordered
counterpart, Theorem \ref{7.1theo} and Corollary \ref{7.1cor}
remain true if one restricts the sums on the right hand side of
(\ref{7.5eqa})and (\ref{7.6eqa}), respectively, to gen. Feynman
graphs without self-contractions. An extension to the models
described in Theorem \ref{5.2theo} is also straight forward, cf.
the last paragraph of Section \ref{6sec}.
\section{Classical particles in the grand canonical ensemble}
\label{8sec} In this section we apply the results of the two
preceding sections to the models of Section 4. We start with a
summary of the physical interpretation of these models, see also
\cite{AGY1,AGY2}:

Let  in (\ref{4.1eqa}) be $z>0$ and $a,\sigma^2=0$, i.e. $\psi(t)$
is purely Poisson. We again consider the measure $\rho_0$
associated with ${\cal C}_{\rho_0}(h)=\exp\{\int_{\R^d}\psi(h)\,
dx\}$, $h\in{\cal S}$. The coordinate process $\eta$ is a marked Poisson process
with intensity $z$, where the mark space is $\R$ and the
distribution of marks $r$. In other words, $\eta$ has the
interpretation of noninteracting classical, continuous particles
in the configurational grand canonical ensemble with activity $z$,
see e.g. \cite{Ru}, where each particle carries a $r$-distributed
random charge with $r$ as in (\ref{4.1eqa}). The random field
$\phi=g*\eta$ obtained as the solution of $L\phi=\eta$ then has
the natural interpretation as a static (short range) field
associated to the charge distribution $\eta$.  The interaction of
the system of charged particles $\eta$ can then be defined as
$U_\Lambda(\eta)=V_\Lambda(g*\eta)$ with
$V_\Lambda(\phi)=\int_\Lambda v(\phi)\, dx$ where we have tacitly
uv-regularized the kernel $g=g_\epsilon$ which implies that the
random field $\phi$ has continuous paths, or, equivalently that
$\nu_0=\nu_0^\epsilon$, the probability measure associated with
$\phi$, fulfills property 1 of Section 2.

The grand canonical partition function  is defined as
$\Xi_\Lambda=\langle e^{-\beta U_\Lambda}\rangle_{\rho_0}$ with
$\beta={1\over k_BT}$ the inverse temperature, $k_B$ is Boltzmann's
constant. Note that $\nu_0$ is the image measure of $\rho_0$ under
the mapping ${\cal S}'\ni\eta\to \phi=g*\eta\in{\cal S}'$. By the
transformation formula of measures
\begin{equation}
\label{8.1eqa}
\Xi_\Lambda=\langle e^{-\beta U_\Lambda}\rangle_{\rho_0}=\langle e^{-\beta V_\Lambda}\rangle_{\nu_0}=Z_\Lambda.
\end{equation}
Hence, for $v(\phi)=\sum_{p=0}^{\bar p}\lambda_p\,\phi^p$, the expansion obtained in Theorem \ref{6.2theo} in combination with the Feynman rules Theorem \ref{4.1theo} is valid for $\beta p(\beta,z)=\lim_{\Lambda\nearrow\R^d} \log \Xi_\Lambda(\beta)/|\Lambda|$ where $p(\beta,z)$ is the pressure function, cf. \cite[Theorem 3.4.6.]{Ru}.

Furthermore, let $d\rho_\Lambda(\eta)=\Xi_\Lambda^{-1}e^{-\beta U_\Lambda(\eta)}d\rho_0(\eta)$ be the interacting grand canonical measure, then the transformation formula yields
\begin{equation}
\label{8.2eqa}
\langle \eta(x_1)\cdots \eta(x_n)\rangle_{\rho_\Lambda}=\langle (L\phi)(x_1)\cdots (L\phi)(x_n)\rangle_{\nu_\Lambda}=L^{\otimes n}\langle\phi(x_1)\cdots\phi(x_n)\rangle_{\nu_\Lambda}.
\end{equation}
This obviously implies  $\langle \eta(x_1)\cdots
\eta(x_n)\rangle_{\rho_\Lambda}^T=L^{\otimes
n}\langle\phi(x_1)\cdots\phi(x_n)\rangle_{\nu_\Lambda}^T$.
Summarizing the above discussion, we get

\begin{theo}
\label{8.1theo} The expansions obtained  for the (truncated)
moment functions in Theorem \ref{7.1theo} and Corollary
\ref{7.1cor} holds also for the (truncated) moments of the
interacting grand-canonical measure $\rho_\Lambda$ if ${\cal
V}_\Lambda[G]$ defined in Theorem \ref{4.1theo} is modified in the
sense that for an edge connecting an inner empty and an outer full
vertex $\stackrel{x}{
\times}\!\!\!\!\!-\!\!\!-\!\!\!\stackrel{z}{\circ}$ the propagator
function $g(x-z)$ is replaced with a Dirac delta
function\footnote{Even though there is some similarity, the
Feynman rules in this theorem should not be mixed up with the
Feynman rules for the amputated Green's functions in the
calculation of effective actions in the renormalization group
\cite{Go}.} $\delta(x-z)$.

In particular, the TD-limit of the (truncated) moments
$\langle\eta(x_1)\cdots\eta(x_n)\rangle^T_\rho=\lim_{\Lambda\nearrow\R^d}\linebreak\langle\eta(x_1)\cdots\eta(x_n)\rangle_{\rho_\Lambda}^T$ and
$\langle\eta(x_1)\cdots\eta(x_n)\rangle_\rho=\lim_{\Lambda\nearrow\R^d}\langle\eta(x_1)\cdots\eta(x_n)\rangle_{\rho_\Lambda}$
exists in the sense of formal power series. Furthermore, the pressure function $p(\beta,z)$
in the sense of formal power series is given by $k_BT$ times the
right hand side of (\ref{6.7eqa}).
\end{theo}

Let us go one step further and consider the case where in
(\ref{4.1eqa}) $z>0$ and $\sigma^2>0$. There is a Gaussian and a
(marked) Poisson contribution to the random field $\eta$. While
the Poisson contribution is interpreted as grand canonic ensemble
of mesoscopic charged particles, the Gaussian contribution can be
interpreted as a white noise fluctuation of the charge density due
to microscopic particles.\footnote{In fact, the Gaussian part can
be seen as the scaling limit of a Poisson contribution, $\eta_z$,
neutral in average, where the intensity $z\to \infty$ and the
charges are being scaled $\sim 1/\sqrt{z}$. I.e. in (\ref{4.1eqa})
we take $\sigma^2=0,a=0$ and $r$, fulfilling $c_1=0$, is replaced
with $r_z(A)=r(\sqrt{ z}A)$ for $A\subseteq \R^d$ measurable.
Taking the limit $\lim_{z\to\infty}\psi_z(t)=c_2t^2/2$ implies
that $\eta_z$ converges in law to a Gaussian white noise as
$z\to\infty$, cf.  \cite{AGY1,AGY2} for the details.} The random field $\eta$ now
stands for the total random charge distribution containing the
mesosopic and the microscopic part. The above analysis can be
repeated word by word and Theorem \ref{8.1theo} also gives the
expansions of the pressure and the (truncated) moment functions of
the given mixed system containing two clearly separated scales. It
is also clear, that there is Wick-ordered version of Theorem
\ref{8.1theo}.

Having set the frame, we want to do calculations  for some
specific examples, where the diagrammatic structure is
particularly simple. This is e.g. the case, when the measure
$\nu_0$ is symmetric and all inner empty vertices with an odd
number of legs vanish, cf. Corollary \ref{3.2cor}. In the given
situation, this can be achieved choosing the charge distribution
$r$ of the non-interacting gas symmetric, $r(-A)=r(A)$ $\forall
A\subseteq \R$ measurable, and $a=0$ which implies $c_n=0$ for odd
$n\in2\N+1$. Furthermore, the simplest non-trivial kind of
interaction is $v(\phi)=\lambda_2\phi^2$ for $\lambda_2>0$. Here
we do not use Wick-ordering as it is more difficult to interpret
and the only term it removes in the expansion of the free energy
density is the first order contribution, which is easy to
calculate.

To understand this interaction, let
$\eta=\sum_{l=1}^ns_l\delta_{y_l}$ be a finite, discrete  charge
distribution and $U(\eta)=V(g*\eta)=\int_{\R^d}v(\phi)\, dx$ is
the potential energy without cut-offs. One obtains
\begin{equation}
\label{8.3eqa}
U(\eta)=\lambda_2\int_{\R^d}\left(\sum_{l=1}^ns_l\,g(y_l-y)\right)^2 dy=\sum_{j,l=1}^ns_js_l\,\tilde g_1(y_j-y_l)+\sum_{l=1}^ns_l^2\tilde g_1(0),
\end{equation}
where $\tilde g_1=\lambda_2\,g*g$. The second sum on the  right
hand side of (\ref{8.3eqa}) can be seen as self energy term or a
(negative) chemical potential that depends on the charge $s$ of
the particle. It can be removed by an adaptation of $z$ and
$r$.\footnote{Take e.g. the simplest case where
$r=(\delta_c+\delta_{-c})/2$ and $s_l^2\equiv c^2>0$. As $z=(2\pi
M/\beta)^{d/2}e^{\beta\mu}$ with $M>0$ the mass of the particles
(assumed to be equal for particles with positive and negative
charge) and $\mu$ the chemical potential, one can compensate the
self energy term by replacing $\mu$ with $\mu+c^2\tilde g_1(0)$.}
The first sum is a usual pair interaction potential for charged
particles. A $\phi^p$ interaction would also contain $l$-body
potentials for $l\leq p$.

We want to  calculate the free energy density $f=\beta p(z,\beta)$
for small $\beta$ and $z$ (low density high temperature regime).
In the diagrammatic expansion given in Theorem \ref{6.2theo}, only
two-legged interaction vertices appear. Like in Fig. 3 in Section
\ref{4sec}, we can introduce a new type of edge denoted by a thin
line
\begin{picture}(2.5,.6)
\thicklines
\linethickness{.4mm}
\put(.05,.1){\line(1,0){.3}}
\put(.33,-.02){$\bullet$}
\put(.5,.1){\line(1,0){.3}}
\put(1.1,-.02){$=$}
\thinlines
\put(1.6,.1){\line(1,0){.7}}
\put(.1,.25){$g$}
\put(.6,.25){$g$}
\put(1.85,.25){$\tilde g_1$}
\end{picture}
and we get  that the gen. Feynman graphs of $m$-th order are
exactly all graphs with an arbitrary number of indistinguishable
inner empty vertices with an arbitrary number of indistinguishable
legs
 and exactly $m$ "thin" edges connecting two inner empty vertices. We note that, as the legs of
\begin{picture}(1,.3)
\thicklines
\linethickness{.4mm}
\put(.05,.1){\line(1,0){.3}}
\put(.33,-.015){$\bullet$}
\put(.47,.1){\line(1,0){.3}}
\end{picture}
are distinguishable,  the thin edge has to be treated as a
directed edge in order to get the right multiplicity of a given
graph. The evaluation rules ${\cal V}'[G]$ for a graph of this new
type are simply to replace each thin edge by $\tilde g_1$ and to
multiply with $c_n$ for each inner empty vertex with $n$ legs.
Then one integrates over all but one of the inner empty vertices.
That this description in fact gives the right rules, i.e. that the
infra-red cut-off $\Lambda$ in the TD limit can be shifted from
the integration over the inner full vertices to the inner empty
vertices, follows from the argument of Appendix \ref{Bsec}.

Figure 6 shows the  graphs $G$ that are contributing to the free
energy density up to fourth order together with their multiplicity
and value ${\cal V}'[G]$. $\tilde g_n=\tilde g^{*n}_1$ is the
$n$-fold convolution of $\tilde g_1$ with itself.
\begin{figure}
\begin{center}
\begin{picture}(15,14)
\thinlines

\put(1,13.5){$\bigcirc$}
\put(1.36,13.6){\oval(.8,.2)[r]}
\put(.4,13.5){1)}
\put(.4,12.8){$m=1,\,u=1$}
\put(.4,12.3){${\cal V}'=c_2\tilde g_1(0)$}

\put(6.9,13.6){\oval(.8,.2)[l]}
\put(6.85,13.5){$\bigcirc$}
\put(7.2,13.6){\oval(.8,.2)[r]}
\put(5.9,13.5){2)}
\put(5.9,12.8){$m=2,\, u=1$}
\put(5.9,12.3){${\cal V}'=c_4 \tilde g_1(0)^2$}

\put(12,13.5){$\bigcirc$}
\put(13,13.5){$\bigcirc$}
\put(12.35,13.7){\line(1,0){.7}}
\put(12.35,13.5){\line(1,0){.7}}
\put(11.4,13.5){3)}
\put(11.4,12.8){$m=2,\,u=2$}
\put(11.4,12.3){${\cal V}'=c_2^2\tilde g_2(0)$}

\put(1.55,10.6){\oval(.8,.2)[l]}
\put(1.5,10.5){$\bigcirc$}
\put(1.86,10.6){\oval(.8,.2)[r]}
\put(1.71,10.75){\oval(.2,.8)[t]}
\put(.4,10.5){4)}
\put(.4,9.8){$m=3,\,u=1$}
\put(.4,9.3){${\cal V}'=c_6\tilde g_1(0)^3$}

\put(6.5,10.5){$\bigcirc$}
\put(7.5,10.5){$\bigcirc$}
\put(6.85,10.7){\line(1,0){.7}}
\put(6.85,10.5){\line(1,0){.7}}
\put(7.85,10.6){\oval(.8,.2)[r]}
\put(5.9,10.5){5)}
\put(5.9,9.8){$m=3,\,u=12$}
\put(5.9,9.3){${\cal V}'=c_4c_2\tilde g_2(0)\tilde g_1(0)$}

\put(12,10.5){$\bigcirc$}
\put(13,10.5){$\bigcirc$}
\put(12.5,11.2){$\bigcirc$}
\put(12.27,10.77){\line(4,5){.31}}
\put(12.8,11.15){\line(4,-5){.31}}
\put(12.37,10.6){\line(1,0){.65}}
\put(11.4,10.5){6)}
\put(11.4,9.8){$m=3,\,u=8$}
\put(11.4,9.3){${\cal V}'=c_2^3\tilde g_3(0)$}

\put(1.55,7.75){\oval(.8,.2)[l]}
\put(1.5,7.65){$\bigcirc$}
\put(1.86,7.75){\oval(.8,.2)[r]}
\put(1.71,7.9){\oval(.2,.8)[t]}
\put(1.71,7.6){\oval(.2,.8)[b]}
\put(.4,7.5){7)}
\put(.4,6.8){$m=4,\,u=1$}
\put(.4,6.3){${\cal V}'=c_8\tilde g_1(0)^4$}

\put(6.9,7.5){$\bigcirc$}
\put(7.9,7.5){$\bigcirc$}
\put(7.25,7.7){\line(1,0){.7}}
\put(7.25,7.5){\line(1,0){.7}}
\put(8.25,7.6){\oval(.8,.2)[r]}
\put(6.93,7.6){\oval(.8,.2)[l]}
\put(5.9,7.5){8)}
\put(5.9,6.8){$m=4,\,u=24$}
\put(5.9,6.3){${\cal V}'=c_4^2\tilde g_2(0)\tilde g_1(0)^2$}

\put(12,7.5){$\bigcirc$}
\put(13,7.5){$\bigcirc$}
\put(12.35,7.7){\line(1,0){.7}}
\put(12.35,7.5){\line(1,0){.7}}
\put(13.35,7.6){\oval(.8,.2)[r]}
\put(13.2,7.75){\oval(.2,.8)[t]}
\put(11.4,7.5){9)}
\put(11.4,6.8){$m=4,\,u=24$}
\put(11.4,6.3){${\cal V}'=c_6c_2\tilde g_2(0)\tilde g_1(0)^2$}

\put(1,4.5){$\bigcirc$}
\put(2,4.5){$\bigcirc$}
\put(1.3,4.75){\line(1,0){.8}}
\put(1.36,4.65){\line(1,0){.68}}
\put(1.36,4.55){\line(1,0){.68}}
\put(1.3,4.45){\line(1,0){.8}}
\put(.4,4.5){10)}
\put(.4,3.8){$m=4,\,u=8$}
\put(.4,3.3){${\cal V}'=c_4^2\int_{\R^d}\tilde g_1^4\,dx$}

\put(6.6,4.5){$\bigcirc$}
\put(7.3,4.5){$\bigcirc$}
\put(8,4.5){$\bigcirc$}
\put(6.95,4.7){\line(1,0){.4}}
\put(7.65,4.5){\line(1,0){.4}}
\put(7.65,4.7){\line(1,0){.4}}
\put(6.95,4.5){\line(1,0){.4}}
\put(5.9,4.5){11)}
\put(5.9,3.8){$m=4,\,u=48$}
\put(5.9,3.3){${\cal V}'=c_4c_2^2\tilde g_2(0)^2$}

\put(12,4.5){$\bigcirc$}
\put(13,4.5){$\bigcirc$}
\put(12.5,5.2){$\bigcirc$}
\put(12.27,4.77){\line(4,5){.31}}
\put(12.8,5.15){\line(4,-5){.31}}
\put(12.37,4.6){\line(1,0){.65}}
\put(13.35,4.6){\oval(.8,.2)[r]}
\put(11.4,4.5){12)}
\put(11.4,3.8){$m=4,\,u=96$}
\put(11.4,3.3){${\cal V}'=c_2^2c_4\tilde g_3(0)\tilde g_1(0)$}

\put(6.6,1.5){$\bigcirc$}
\put(7.5,1.5){$\bigcirc$}
\put(6.6,2.2){$\bigcirc$}
\put(7.5,2.2){$\bigcirc$}
\put(6.99,1.6){\line(1,0){.51}}
\put(6.99,2.3){\line(1,0){.51}}
\put(6.8,1.8){\line(0,1){.31}}
\put(7.7,1.8){\line(0,1){.31}}
\put(5.9,1.5){13)}
\put(5.9,.8){$m=4,\,u=48$}
\put(5.9,.3){${\cal V}'=c_2^4\tilde g_4(0)$}

\end{picture}
\end{center}
\caption{\footnotesize Graphs for the gas of charged  particles,
neutral in average, with pair interaction up to fourth order.
$m=$order, $u=$multiplicity, ${\cal V}'$=value. }
\end{figure}
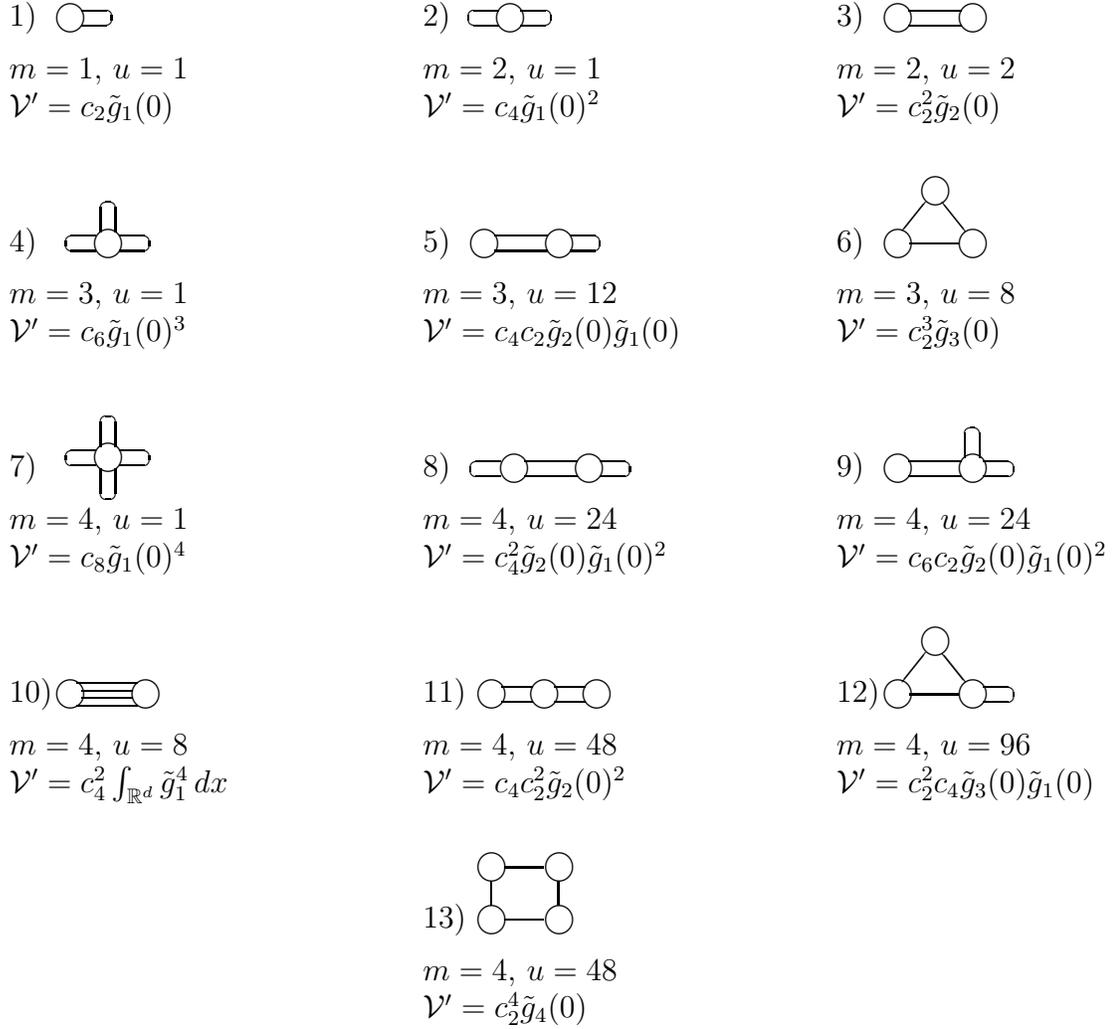

Let us consider a  simple example in $d=2$ dimensions with only
two kinds of charge $\pm c$, i.e.
$c_n=\delta_{2,n}\sigma^2+c^nz/2$, $\sigma$ being the intensity of
the Gaussian background, and $L=(-\Delta+m^2_0)$ with $m_0$ the
range of the interaction.  We get that
$g(x)=(2\pi)^{-2}\int_{\R^2}{1\over (|k|^2+m_0^2)}\linebreak
\times e^{ik\cdot x}\,dk$ diverges logarithmically at $0$. The
measure $\nu_0$ thus does not fulfill property 1 of Section 2. In
fact, the Poisson contribution to the  random field $\eta$
associated to $\rho_0$ has discrete support and the random field
$\phi=g*\eta$ has singularities on the support of the Poisson part
of $\eta$. However, by choosing the interaction to be $\phi^2$, we
see that only $\tilde g_1=\lambda_2\,g*g$ enters into the
perturbation series which is a continuous function for $d<4$. The
uv-cut-offs therefore can be removed from the perturbation series.

The perturbation  coefficients up to forth order can now be
calculated explicitly by solution of rather elementary integrals:
\begin{equation}
\label{8.4eqa}
\tilde g_n(0)=\lambda_2^n(2\pi)^{-1}\int_0^\infty {a\,da\over (a^2+m_0^2)^{2n}}=\lambda_2^n{m_0^{2-4n}\over 2\pi(4n-2)}\end{equation}
and
\begin{equation}
\label{8.5eqa}
\int_{\R^d}\tilde g_1^4dx=\lambda_2^4(2\pi)^{-6}\int_{\R^2}\left[\int_{\R^2}{1\over ((|k-q|^2+m_0^2)^2(|q|^2+m_0^2)^2}\, dq\right]^2dk=\lambda_2^4{m_0^{-6}\over 64 \pi^3 }.
\end{equation}
This gives the following equation of state:
\begin{eqnarray}
\label{8.6eqa}
p(z,\sigma,c,\beta,\lambda_2,m_0)&=&-\lambda_2{\sigma^2+{z\over 2}c^2\over 4m_0^2\pi}+{\lambda_2^2\beta\over 2}\left({zc^4\over 32m_0^4\pi^2}+{(\sigma^2+{z\over 2}c^2)^2\over 6m_0^6\pi}\right)\nonumber\\
&-&{\lambda_2^3\beta^2\over 6}\left( {zc^6\over 128 m_0^6\pi^3}+{(s^2+{z\over 2}c^2)zc^4\over 8 m_0^8\pi^2}+{2(\sigma^2+{z\over 2}c^2)^3\over 5m_0^{10}\pi}\right)\\
&+&{\lambda_2^4\beta^3\over 24}\left( {zc^8\over 512 m_0^8\pi^4}+{z^2c^8\over 32 m_0^{10}\pi^3}+{(\sigma^2+{z\over 2}c^2)zc^6\over 16 m_0^{10}\pi^3}+{z^2c^8\over 32m_0^6\pi^3}\right.\nonumber\\
&+& \left.{(\sigma^2+{z\over 2}c^2)^2zc^4\over 6 m_0^{12}\pi^2}+{3(\sigma^2+{z\over 2}c^2)^2zc^4\over 10 m_0^{12}\pi^2}+{12(\sigma^2+{z\over 2}c^2)^4\over 7m_0^{14}\pi}\right)+{\cal O}(\lambda_2^5).\nonumber
\end{eqnarray}
\begin{rem}
\label{8.1rem} {\rm The graphs that  one obtains for the
$\phi^2$-interaction are obviously very similar to those of the
Mayer series \cite{Re,Ru,TKS}. If there is only one type of
particles with charge $c$, the main difference (neglecting
combinatorial matters) is that in the Mayer series edges are
evaluated with the Mayer function $w(y_1-y_2)=e^{-\beta c^2 \tilde
g_1(y_1-y_2)}-1$ instead of $\tilde g_1(y_1-y_2)$, which of course
is a big advantage if the two point potential $\tilde g_1$ has a
(repulsive) singularity at zero as e.g. in the case of the
Lennard-Jones potential. In such cases, the perturbation expansion
given in this article becomes plagued by very non-trivial
uv-singularities but the Mayer series is not. This is the reason
why the perturbetion expansion, though in principle known to
physicists, see e.g. \cite[Sect. 3.3, Eq. 42]{TKS}, is not
particularly popular. The Mayer series for the gas of particles
with two types of charges $\pm c$ however also contains the Mayer
function $w(y_1-y_2,+,-)= e^{+\beta c^2 \tilde g_1(y_1-y_2)}-1$
for the interaction of a $+$ charge with a $-$ charge which is
more singular than $\tilde g_1(x-y)$. Also, the analytic
calculation in low orders of the graphs including "propagators"
$\tilde g_1(y_1-y_2)$ seems to be more easy than for the
propagators $w(y_1-y_2)$.  In some particular situations, there
might therefore be some physical interest in the derived series
expansion, even though it is not the objective of this article to
solve a concrete physical problem in the thermodynamics of gases,
fluids or electrolytes. }\prend
\end{rem}
At the end of this section, we want to  give some brief and non
technical remarks on the uv-problem, leaving most of the work for
the future. If the measure $\nu_0$ from Section \ref{4sec} is not
Gaussian and $\tilde g_1(y)$ for $y\to0$ has an algebraic
singularity $\sim|y|^{-\theta}$, $\theta>0$, already the
$\phi^2$-perturbation series is not power-counting renormalizable:
If one e.g. considers graphs of the kind 3) and 10) in Fig. 6 for
an arbitrary (even) number $m$ of legs, one gets the suspicious
degree of divergence $m\theta-d$.

Nevertheless,  if one takes a look (\ref{8.3eqa}) in the special
situation where there is only one type of particle with charge
$c>0$, one can see that a simple 1st order local counterterm
$\lambda_1^\epsilon\phi$ (i.e. a chemical potential) with
$\lambda_1=- cg^\epsilon_1(0)/\int_{\R^d}g\, dx$ removes all
singularities in the limit $\epsilon \searrow 0$. This is in
striking contrast with the non-renormalizability.

We say that a  graph has a self-contraction of the second kind if
a subgraph
\begin{picture}(.8,.3)
\thinlines
\put(.1,0){$\bigcirc$}\put(.44,.1){\oval(.6,.2)[r]}
\end{picture}
occurs, cf. the  graphs 1), 2), 4), 5), 7), 8), 9) and 12) of Fig.
6.  If one includes the above counter term into the perturbation
series for the system with only one particle species, one can
prove that all self-contractions of 2nd kind are being removed
from the series. In fact, for each such graph, there is exactly
one other graph where the self-contraction of the 2nd kind is
replaced by
\begin{picture}(1,.3)
\thicklines
\linethickness{.4mm}
\put(.1,0){$\bigcirc$}\put(.46,.1){\line(1,0){.39}}\put(.8,.1){\circle{.2}}\put(.8,.1){\circle*{.09}}
\end{picture},
 with
\begin{picture}(.5,.3)
\thicklines
\put(.2,.1){\circle{.2}}\put(.2,.1){\circle*{.09}}
\end{picture}
the interaction vertex of the linear counterterm. It is therefore
clear that self-contractions of 2nd kind are caused by the
self-energy terms on the right hand side of (\ref{8.3eqa}).

This observation has  two immediate consequences: Firstly, in the
case where $\tilde g_1(y)$ has an algebraic singularity at $0$,
the perturbation series remains non-power counting renormalizable,
even though it can be "summed up" and then gives a finite result
\cite{Ru}.

Secondly, if the  singularity of $\tilde g_1(y)$ at $y=0$ is only
logarithmic, the self-contractions of 2nd kind are the only source
of divergences\footnote{The situation has some similarity with
Gaussian $\phi^4$-theory in $d=3$ dimensions (take
$L=(-\Delta+m_0^2)^{1/2}$), where there is also just one subgraph
(see Fig. 3) causing logarithmic divergences.}, cf. the proof of
Theorem \ref{5.2theo} for the uv-finiteness of
2nd-self-contraction free graphs. The given choice of the
counterterm removes the uv-divergences from the perturbation
series.  This e.g. occurs in the cases $d=4$ and
$L=(-\Delta+m_0^2)$, $d=2$ and $L=(-\Delta+m_0^2)^{1/2}$ or $d=2$,
$L=(-\Delta+m_0^2)$ and
$V_\Lambda(\phi)=\lambda_2\int_\Lambda|\nabla\phi|^2\, dx$ is of
gradient type leading to a pair potential $\tilde
g_1(x)=(2\pi)^{-2}\int_{\R^2}{|k|^2\over (|k|^2+m^2)^2}\,
e^{ik\cdot x}\, dk$ with equally strong repulsive and attractive
parts, i.e. $\int_{\R^2} \tilde g_1\,dx=0$.

\appendix

\section{Facts about truncation}
\label{Asec} For the convenience of the reader, we give proofs of
the well-known facts F1) -- F4) on the combinatorics of truncation
starting with F4):
\begin{lem}
\label{A.1lem} Let $u : C\to \C$ for $C$ an open set in ${\cal S}$
be infinitely often partial differentiable and let $w : \C \mapsto
\C $ be analytic  on an open neighborhood of $u(C)$. Then for
$\{h_n\}_{n\in\N} \subseteq {\cal S}$ and $J\subseteq \N$ finite
\begin{equation}
\label{A.1eqa}
\partial_{J} w \circ u = \sum^{\sharp J}_{k = 1}
w^{(k)} \circ u \sum_{I \in {\cal P}(J)\atop I=\{I_1,\ldots,I_k\}}
\prod_{l=1}^k \left[\partial_{I_l} u\right]
\end{equation}
holds on $C$.
Here $ \partial_{A}=\left[\prod_{j\in A}{\partial\over\partial h_j}\right]$ for $A\subseteq \N$ finite and $w^{(k)} (z) = \left( {d^k \over dz^k} w \right) (z)$.
\end{lem}
The proof is by use  of Leibnitz' chain rule and induction over
$\sharp J$, details can be found in \cite[Lemma 3.3]{AGW1}. The
application of this generalized chain rule to $J=\{1,\ldots,n\}$,
$u={\cal C}^T$ ($\Rightarrow \, {\cal C}^T(0)=0$) and $w$ the
exponential function, evaluation at $0\in {\cal S}$ and doing the
limit $h_l\to\delta_{x_l}$, $l\in J$, establishes fact F4. We note
that in Lemma \ref{A.1lem} one can also replace ${\cal S}$ with
$\R_+=[0,\infty)$ and do right derivatives at zero for $u$
infinitely often right differentiable, as required in Eq.
(\ref{6.1eqa}). F1) is immediate from F4).

We prove F2) in the  form required in Section \ref{6sec}. Let
$X,Y\subseteq \N$ disjoint, then
\begin{equation}
\label{A.2eqa}
\langle XY\rangle_{\nu_0}-\langle X\rangle_{\nu_0}\langle Y\rangle_{\nu_0}=\sum_{I\in{\cal P}_c(X,Y)\atop I=\{I_1,\ldots, I_k\}}\prod_{l=1}^k\langle I_l\rangle_{\nu_0}^T.
\end{equation}
As $I\in{\cal P}_c(X,Y)$  there exists at least one $l=1,\ldots,k$
such that $X_l=I_l\cap X\not=0$ and $Y_l=I_l\cap Y\not=0$. If the
truncated moment functions vanish exponentially for large
separation of their arguments, we get that each term in the sum on
the right hand side contains at least factor $|\langle
I_l\rangle^T|\leq  D'\exp\{-m_0 \underline{d}(X_l,Y_l)\}\leq
D'\exp\{-m_0 \underline{d}(X,Y)\} $. The right hand side thus
vanishes exponentially for large separation of $X$ and $Y$.

Conversely, let $\nu_0$ have the exponential clustering property.
We proceed by induction over $n=\sharp X+\sharp Y$. If we set
$\langle\emptyset\rangle^T_{\nu_0}=0$ and $d(X,\emptyset)=0$, the
assertion $|\langle XY\rangle^T_{\nu_0}|\leq D'\exp\{-m_0
\underline{d}(X,Y)\}$ is trivial for $n=0$. Suppose that it holds
up to $n-1$, then each term on the right hand side of
(\ref{A.2eqa}) except for the term $I=\{X\cup Y\}$ contains at
least one factor on which the induction hypothesis applies and
which thus vanishes exponentially fast as $X$ and $Y$ get
separated. As the left hand side also vanishes exponentially fast,
this must also apply to this remaining term $I=\{X\cup Y\}$. Hence
F2) holds.

To get F3), consider Eq. (\ref{3.3eqa}).  If $n$ is odd, for each
partition $I=\{I_1,\ldots,I_k\}$ on the right hand side there
exists at least one $l\in\{1,\ldots,k\}$ such that $\sharp I_l$ is
odd. Hence the vanishing of $\langle J\rangle ^T_{\nu_0}$ for
$J\subseteq \N$ with $\sharp J$ odd implies the vanishing of the
left hand side of (\ref{3.3eqa}).

Let conversely  the odd moments of $\nu_0$ be vanishing. We
proceed by induction over $l$ and let $\sharp X=2l+1$. For $l=0$
we get $\langle X\rangle_{\nu_0}=\langle X\rangle^T_{\nu_0}=0$.
Suppose that $\langle X\rangle_{\nu_0}^T=0$ for odd $\sharp
X<2l+1=n$. Hence all term on the right hand side of (\ref{3.3eqa})
except for the one with $I=\{\{1,\ldots,n\}\}$ vanish. But the
left hand side is zero, and this remaining term therefore must be
zero, too.

\section{TD limit for certain integrals}
\label{Bsec}
Let $I(y_1,\ldots,y_m)$ be a translation invariant function such that
\begin{eqnarray}
\label{B.0eqa}
|I(y_1,y_2,\ldots,y_m)|&\leq& C\left( 1+B\sum_{l,j=1\atop l\not=j}^m1_{\{ |y_l-y_j|<1\}}(y_j-y_l)|\log|y_l-y_j||^n\right)\nonumber\\
&&~~~~~~~~~~~~~~~~~~~~~~~~~~~~~~~~~~\times\exp\{-M\sum_{l=2}^m|y_l-y_1|\}
\end{eqnarray}
 for some $n\in\N$ and $M,B,C>0$. $1_A$ stands for the indicator function of the set $A$. Then the following holds in the TD limit:
\begin{equation}
\label{B.1eqa}
\lim_{\Lambda\nearrow\R^d}{1\over|\Lambda|}\int_{\Lambda^m}I(y_1,\ldots,y_m)\, dy_1\cdots dy_m=\int_{\R^{d(m-1)}}I(0,y_2,\ldots,y_m)\,dy_2\cdots dy_m.
\end{equation}
In fact, by Fubini's theorem and translation invariance
\begin{equation}
\label{B.2eqa}
{1\over|\Lambda|}\int_{\Lambda^m}I(y_1,\ldots,y_m)\,dy_1\cdots dy_m ={1\over |\Lambda|}\int_{\Lambda_{y_1}}\left[\int_{\Lambda_{y_1}^{m-1}}I(0,y_2,\ldots,y_m)\, dy_2\cdots dy_m\right]dy_1,
\end{equation}
where $\Lambda_{y_1}=\Lambda-y_1$. We consider the expression in the brackets $[\cdots]$ as a function of $y_1$ and we obtain
\begin{eqnarray}
\label{B.3eqa}
[\cdots]&=&\int_{\R^{d(m-1)}}I(0,y_2,\ldots,y_m)\,dy_1\cdots dy_m\nonumber\\
&-&\sum_{l=0}^{m-2}(-1)^{l+1}\int_{\R^{dl}}\int_{(\R^d\setminus \Lambda_{y_1})}\int_{\Lambda_{y_1}^{m-2-l}} I(0,y_2,\ldots,y_m)\, dy_2\cdots dy_m .
\end{eqnarray}
We want to get an estimate for the sum on the right hand side of (\ref{B.3eqa}): Using (\ref{B.0eqa}) one obtains
\begin{eqnarray}
\label{B.4eqa}
&&\left|\sum_{l=0}^{m-2}(-1)^{l+1}\int_{\R^{dl}}\int_{(\R^d\setminus \Lambda_{y_1})}\int_{\Lambda_{y_1}^{m-2-l}} I(0,y_2,\ldots,y_m)\, dy_2\cdots dy_m \right|\nonumber\\
&\leq& C(m-1) \left(\int_{\R^d} e^{-M|y|}dy+Bm\int_{\{|y|\leq 1\}}|\log|y||^n\, dy\right)^{m-2} u(y_1,\Lambda)
\end{eqnarray}
where $u(y_1,\Lambda)=\int_{\R^d\setminus\Lambda}e^{-M|y-y_1|}(1+B1_{\{|y-y_1|<1\}}|\log|y-y_1||^n)\,dy$. Hence, for $C'>0$ large enough,
\begin{eqnarray}
\label{B.5eqa}
&&\left|{1\over|\Lambda|}\int_{\Lambda^m}I(y_1,\ldots,y_m)\,dy_1\cdots dy_m\right.\nonumber\\
&&\left.~~~~~~-\int_{\R^{d(m-1)}}I(0,y_2,\ldots,y_m)\,dy_2\cdots dy_m\right|\leq C'{1\over |\Lambda|}\int_{\Lambda} u(y_1,\Lambda)\, dy_1
\end{eqnarray}
We note that $u(y_1,\Lambda)\leq C'' e^{-{M\over 2}\underline{d}(\partial\Lambda,y_1)}$  for $y_1\in
\Lambda$ where $\underline{d}(\partial\Lambda,y_1)$ stands for the distance from $y_1$ to the boundary of $\Lambda$ and $C''=\int_{\R^d}e^{-{M\over 2}|y|}(1+B|\log|y||^n)\,dy$. Let $\partial_a\Lambda=\{ y\in\Lambda: \underline{d}(\partial\Lambda,y_1)<a\}$,
then
\begin{equation}
\label{B.6eqa}
{1\over|\Lambda|}\int_\Lambda u(y_1,\Lambda)\, dy_1\leq {C''\over|\Lambda|}\left(e^{-Ma/2}|\Lambda\setminus\partial_a \Lambda|+|\partial_a\Lambda|\right)\leq C''(e^{-Ma/2}+|\partial_a\Lambda|/|\Lambda|).
\end{equation}
holds for all $a>0$. As convergence $\Lambda\nearrow\R^d$ in the sense of Van Hove means that
$|\partial_a\Lambda|/|\Lambda|\to 0$ $\forall a>0$ in the TD limit, the right hand side of (\ref{B.6eqa}) and hence (\ref{B.5eqa}) can be made arbitrarily small  for $\Lambda$ in the TD limit sufficiently large.  This proves
equation (\ref{B.1eqa}).

\vspace{1cm} \noindent{\bf Acknowledgements.} H. G. has been
financially  supported by the D.F.G. through the project
"Stochastic methods in quantum field theory". He also would like
to thank the D\'epartment des Math\'ematiques at Tunis El Manar
for its warm hospitality on repeated occasions.
S.H.D. and H. O. would like to thank Sergio Albeverio for his kind invitation to Bonn through
SFB 611 and D.F.G. project "Systems with infinitely many degrees of freedom".

\vspace{1cm}
\noindent {\sc Sidi Hamidou Djah} and {\sc Habib Ouerdiane}\\
{\small D\'epartement des Mat\'ematiques\\
Universit\'e Tunis El Manar\\
Campus Universitaire,
Tunis 1060, Tunisia\\
{\tt habib.ouerdiane@fst.rnu.tn} and {\tt jah.sidi@fst.rnu.tn}
}

\vspace{.5cm}

\noindent {\sc Hanno Gottschalk}\footnote{Corresponding author} \\
{\small Institut f\"ur angewandte Mathematik\\
Rheinische Friedrich-Wilhelms-Universit\"at Bonn\\
Wegelerstr. 6, D-53115 Bonn, Germany\\
{\tt gottscha@wiener.iam.uni-bonn.de}}
 \end{document}